\documentclass[aps,pre,
preprint,
preprintnumbers,
tightenlines,
superscriptaddress,
amsmath,
amssymb,
noshowkeys,
nofootinbib,
floatfix,
longbibliography,
]{revtex4-2}
\makeatletter
\def\frontmatter@abstract@produce{%
  \par
  \addvspace{\frontmatter@preabstractspace}%
   \begingroup
    \dimen@\baselineskip
    \setbox\z@\vtop{\unvcopy\absbox}%
    \advance\dimen@-\ht\z@\advance\dimen@-\prevdepth
    \@ifdim{\dimen@>\z@}{\vskip\dimen@}{}%
   \endgroup
   \begingroup
    \prep@absbox
    \unvbox\absbox
    \post@absbox
   \endgroup
  \@ifx{\@empty\mini@notes}{}{\mini@notes\par}%
  \addvspace\frontmatter@postabstractspace
}%
\makeatother
\usepackage{graphicx}          % Include figure files
\usepackage{subfigure}         % subfigure package
\usepackage{dcolumn}           % Align table columns on decimal point
\usepackage{capt-of}           % creats captioning of tables and figures
\usepackage{makecell}          % override cells in tables
\usepackage{bbm}               % bold math
\usepackage{bm}                % bold math
\usepackage{time}              % time of day
\usepackage[mathlines]{lineno} % linenumbers in pdf file
\usepackage{stackrel}          % create stacking symbols
\usepackage{enumitem}          % itemize package
\usepackage[mathscr]{eucal}    % Eular script Eular script
\usepackage{xcolor}            % needed for mixed colors (blue!60!black, etc.)
\usepackage{hyperref}
\hypersetup{
  colorlinks = true,
  linkcolor  = blue!60!black,
  citecolor  = green!50!black,
  urlcolor   = blue!60!black,
  pdftitle   = {NLDE with SS, VV, and PP interactions},
  pdfauthor  = {John Dawson}
}
\usepackage{amsthm}
\theoremstyle{plain}                           % default

\theoremstyle{definition}

\theoremstyle{remark}

\newcommand{\rmi}{{\rm i}}                     % roman i
\newcommand{\rmd}{{\rm d}}                     % roman d
\newcommand{\rme}{{\rm e}}                     % roman e
\newcommand{\Partial}[4]%                      % partial derivs w constants
   {\Bigl ( \frac{\partial #1 }{\partial #2 } \Bigr )_{\! #3, #4 }}
\newcommand{\calC}{\mathcal{C}}
\newcommand{\calK}{\mathcal{K}}

\newcommand{\calJ}{\mathcal{J}}
\newcommand{\calL}{\mathcal{L}}

\newcommand{\calN}{\mathcal{N}}

\newcommand{\calT}{\mathcal{T}}
\newcommand{\calV}{\mathcal{V}}
\newcommand{\bbR}{\mathbbm{R}}
\newcommand{\bbC}{\mathbbm{C}}

\newcommand{\bbZ}{\mathbbm{Z}}
\newcommand{\Tproduct}[1]%
   {\ensuremath{\mathcal{T} \{ \, #1 \, \} } }
\usepackage{mathtools}           % loads and fixes amsmath first
\usepackage{amssymb}             % math symbol fonts
\usepackage{mathrsfs}            % elegant script font
\usepackage{derivative}          % derivatives
\usepackage{braket}              % Dirac braket
\usepackage{float}               % [H] float placement (Fig.~GNKappaCVsKappaMax3D)
\usepackage{enumitem}             % [noitemsep] compact enumerate lists
\newcommand{\be}{\begin{equation}}
\newcommand{\ee}{\end{equation}}
\newcommand{\bea}{\begin{eqnarray}}
\newcommand{\eea}{\end{eqnarray}}
\newcommand{\bq}{\begin{equation}}
\newcommand{\ba}{\begin{eqnarray}}
\newcommand{\eq}{\end{equation}}
\DeclareMathOperator{\sgn}{sgn}
\begin{document}
%
%\preprint{LAUR-26XX}
%
\title{Solitary waves in a NLDE with Scalar-Scalar and 
Pseudoscalar-Pseudoscalar Interactions}
\author{John F. Dawson}
\email[Email: ]{john.dawson@mac.com}
\affiliation{
   Department of Physics,
   University of New Hampshire,
   Durham, NH 03824, USA}
\author{Fred Cooper}
\email[Email: ]{cooper@santafe.edu}
\affiliation{
   Santa Fe Institute,
   1399 Hyde Park Road,
   Santa Fe, NM 87501, USA}
\affiliation{
   Center for Nonlinear Studies and Theoretical Division,
   Los Alamos National Laboratory,
   Los Alamos, NM 87545, USA}
\date{\today, \now \ PDT}
%
% -----------------------------------------------------------------------
\begin{abstract}
We study solitary-wave solutions of the nonlinear Dirac equation in
$1$+$1$ dimensions with scalar-scalar (SS), vector-vector (VV), and
pseudoscalar-pseudoscalar (PP) self-interactions of arbitrary
nonlinearity $\kappa$, parametrized by couplings $g_s=g/q$, $g_v=g/p$,
$g_p=g/r$. We derive the closed-form soliton solution, its reality
domain, and its single- versus multi-humped structure for general
couplings, then specialize to two complementary limits, the SS-PP
model ($g_v=0$) and the VV-PP model ($g_s=0$), evaluating the charge
and energy and the domain of convergence in closed form for each. The
Vakhitov-Kolokolov criterion is only a necessary condition for linear
stability, so we go on to diagonalize the Bogoliubov-de~Gennes
eigenvalue problem directly, using a Chebyshev spectral discretization
cross-validated against a SLAC-derivative lattice method and against
direct nonlinear time evolution of the perturbed soliton. For
representative three-hump VV-PP examples we confirm linear stability
beyond the VK criterion, resolving for these cases a question left
open by prior nonlinear-evolution studies of multi-humped solitary
waves. For unstable SS-PP and VV-PP examples we extract the genuine
growing eigenmode in each case and confirm its predicted growth rate
by nonlinear evolution. Beyond the usual translation and phase zero
modes, we also find a discrete, purely oscillatory bound state inside
the spectral gap in some of these cases, distinct from the growing
instability. We discuss the relation of this work to a companion
paper that applies a similar doubler-free lattice method to the pure
SS and VV models without the PP term.
\end{abstract}
\maketitle
%
% -----------------------------------------------------------------------
%
\section{\label{s:Intro}Introduction}

Solitary-wave solutions of the nonlinear Dirac equation (NLDE) in
$1$+$1$ dimensions go back to Soler's scalar-scalar (SS) ansatz
\cite{Soler-1970} and to Thirring's exactly solvable vector-vector
(VV) model \cite{Thirring-1958}, and both families persist for
arbitrary nonlinearity $\kappa$ \cite{PhysRevE.82.036604}. Orbital
stability is settled at the integrable Thirring point itself
($\kappa=1$, VV) \cite{Pelinovsky-Shimabukuro-2014,
Contreras-Pelinovsky-Shimabukuro-2016}, but away from $\kappa=1$ the
picture is far less complete, and what is known is scattered across
specific regimes rather than organized into a single diagram. Evans-function
and related spectral methods settle the cubic ($\kappa=1$) SS case
\cite{Berkolaiko-Comech-2012,BOUSSAID2019108289} and its Gross-Neveu
analogue \cite{Comech-Phan-Stefanov-2017}; the Vakhitov-Kolokolov (VK)
criterion, together with an energy-vanishing condition, identifies
when a real instability eigenvalue can appear at all
\cite{vk,Berkolaiko_2015}; and a direct calculation finds such an
instability setting in near $\omega_0=m$ once $\kappa>2$
\cite{Comech-2014}, with a thorough account of this partial picture
given in Ref.~\cite{BoussaidComech-book-2019}. Separately, nonlinear
time evolution suggests that every double-humped SS soliton is
unstable \cite{PhysRevE.90.032915}, though whether that instability is
already present at the linear level has remained an open question, and
neither the SS nor the VV model has had a complete stability diagram
across the full $(\kappa,\omega_0)$ plane.

The same question is being pursued independently, and over a wider
parameter range, by Cooper, Jacobs, and Khare
\cite{Cooper-Jacobs-Khare-2026}, for the pure SS and pure VV models.
Rather than a closed-form soliton, they work directly from a lattice
formulation of the Bogoliubov-de~Gennes problem, which runs into the
same spurious-eigenvalue obstacle that complicates any naive
discretization of a linearized Dirac operator: fermion doubling,
where a local finite-difference stencil's dispersion relation folds
back on itself at the Brillouin-zone edge and manufactures an
unphysical low-energy mode that does not go away under refinement.
They remove it with the doubler-free SLAC derivative
\cite{Drell-Weinstein-Yankielowicz-1976}, using a box-size scan to
separate genuine, box-independent eigenvalues from finite-volume
artifacts, and apply that combination across the full $(\kappa,\omega)$
plane. What they find: a real instability eigenvalue tracks the sign
of $dQ_0/d\omega_0$ exactly, appearing only for $\kappa>2$ and only
above a $\kappa$-dependent threshold $\omega_c(\kappa)$; no
oscillatory instability ever appears for $\omega>0$ in either model,
so a soliton's hump count by itself says nothing about its stability;
and the VV model's negative-frequency branch, which has no SS
analogue, carries two further instability mechanisms found nowhere
else -- a real eigenvalue exactly where the energy changes sign, and
a genuinely complex quartet straddling the massive Thirring point.

The present paper takes a complementary route to much the same
question, for a model their analysis does not cover: in addition to
the SS and VV terms, we keep a pseudoscalar-pseudoscalar (PP)
self-interaction throughout, parametrizing all three couplings
together as $g_s=g/q$, $g_v=g/p$, $g_p=g/r$ following Cooper, Khare,
Mihaila, and Saxena \cite{PhysRevE.82.036604}, with the $g_p=0$
two-coupling special case studied in detail by Khare, Cooper, Dawson,
and Saxena \cite{Khare-Cooper-Dawson-Saxena-2026}
[Eq.~\eqref{e:chiralGN}]. We keep all three couplings general for as
long as possible -- the equation of motion, the closed-form soliton,
its reality domain, hump structure, and the Vakhitov-Kolokolov
criterion are all derived for arbitrary $(q,p,r)$ -- before
specializing where a particular limit gives a qualitatively simpler or
sharper result. Two such limits receive the most attention here: the
SS-PP model ($g_v=0$) and its complement, the VV-PP model ($g_s=0$),
the latter being where a three-humped numerical example actually lies
(Sec.~\ref{ss:HumpStructure}). Rather than a lattice discretization of the linearized
operator, we diagonalize the Bogoliubov-de~Gennes eigenvalue problem
on a Chebyshev spectral grid mapped onto the whole real line, and
cross-validate the result two ways: against the SLAC-derivative
lattice method of Ref.~\cite{Cooper-Jacobs-Khare-2026} itself, applied
directly to our own soliton and our own growth rates rather than a
separate toy problem, and by evolving the full nonlinear equation
forward in time from a perturbed initial condition and confirming that
the observed growth or oscillation rate matches the
Bogoliubov-de~Gennes prediction. For the specific three-hump VV-PP
examples at $\kappa=1.05$ and $\kappa=3/2$, this direct diagonalization
answers exactly the kind of question left open by nonlinear time
evolution alone \cite{PhysRevE.90.032915}: both are not merely
VK-stable but linearly stable, with every eigenvalue beyond the two
zero modes confirmed real.

A further point of contact: a closely related Vakhitov-Kolokolov
survey, covering a much wider range of couplings and nonlinearities
for the related massive Gross-Neveu model, is given in
Ref.~\cite{Dawson-GN-2026}, which we also draw on for the
Chebyshev-versus-SLAC cross-check of Sec.~\ref{ss:SLACcompare}.

The remainder of the paper is organized as follows.
Section~\ref{s:Model} sets up the Lagrangian, equation of motion, and
conserved charge and energy-momentum tensor for the general
SS+VV+PP model. Section~\ref{s:Soliton} derives the closed-form
soliton solution, its reality domain, and its single- versus
multi-humped structure. Section~\ref{s:ChargeEnergy} evaluates the
charge and energy in closed form and establishes the domain of
convergence. Section~\ref{s:Stability} develops the Vakhitov-Kolokolov
criterion and the Bogoliubov-de~Gennes eigenvalue problem in general,
describes the Chebyshev spectral method used to solve it, and then
specializes to the SS-PP and VV-PP cases in turn, in each case
confirming the result by direct nonlinear time evolution.
Appendix~\ref{s:Notation} collects our notation and conventions.

\section{\label{s:Model}Model Lagrangian}

The Lagrangian density is of the form
\begin{equation}\label{e:DiracEq}
   \calL = \bar{\Psi}( \rmi \gamma^{\mu} \partial_{\mu} -m) \Psi + \calV_{SS} + \calV_{VV} + \calV_{PP} \>,
\end{equation}
where the scalar-scalar (SS), vector-vector (VV), and pseudoscalar-pseudoscalar (PP) interaction terms are given,
for arbitrary nonlinearity parameter $\kappa>0$, by
\begin{subequations}\label{e:VSSVVPP}
\begin{align}
   \calV_{SS} &= \frac{g_s}{(\kappa+1)} \, \bigl[ (\bar\Psi \Psi)(\bar\Psi \Psi) \bigr]^{(\kappa+1)/2} \>,
   \label{e:VSS} \\
   \calV_{VV} &= \frac{g_v}{(\kappa+1)} \, \bigl[ (\bar\Psi \gamma_{\mu} \Psi)(\bar\Psi \gamma^{\mu} \Psi) \bigr]^{(\kappa+1)/2} \>,
   \label{e:VVV} \\
   \calV_{PP} &= \frac{g_p}{(\kappa+1)} \, \bigl[ (\bar\Psi \gamma_{5} \Psi)(\bar\Psi \gamma^{5} \Psi) \bigr]^{(\kappa+1)/2} \>,
   \label{e:VPP}
\end{align}
\end{subequations}
We parametrize the three couplings as
\begin{equation}\label{e:chiralGN}
   g_{s} = g/q
   \qquad
   g_{v} = g/p
   \qquad
   g_{p} = g/r \>, \quad r < 0 \>,
\end{equation}
following Cooper, Khare, Mihaila, and Saxena
\cite{PhysRevE.82.036604}, with the $g_p=0$, two-coupling SS+VV
special case studied in detail by Khare, Cooper, Dawson, and Saxena
\cite{Khare-Cooper-Dawson-Saxena-2026}. We develop the soliton solution
and its reality, hump-structure, convergence, and stability properties
for general couplings first, specializing only where a particular
limit gives a qualitatively simpler or sharper result. Several such
limits are of interest in what follows, most extensively the SS-PP
model obtained from Eq.~\eqref{e:chiralGN} as $p\to\infty$ (i.e.\
$g_v=0$) and the complementary VV-PP model obtained as $q\to\infty$
(i.e.\ $g_s=0$).

%
% -----------------------------------------------------------------------
\subsection{Equation of motion}

Varying Eq.~\eqref{e:DiracEq} with respect to $\bar\Psi$ gives the
equation of motion
\begin{equation}\label{e:EOM}
\begin{split}
   ( \rmi \gamma^{\mu} \partial_{\mu} - m ) \Psi
   &+ g_s \, ( \bar\Psi \Psi )^{\kappa} \, \Psi \\
   &+ g_v \, \bigl[ (\bar\Psi \gamma_{\nu} \Psi)(\bar\Psi \gamma^{\nu} \Psi) \bigr]^{(\kappa-1)/2}
      (\bar\Psi \gamma_{\mu} \Psi) \, \gamma^{\mu} \Psi \\
   &+ g_p \, \bigl[ (\bar\Psi \gamma_{5} \Psi)(\bar\Psi \gamma^{5} \Psi) \bigr]^{(\kappa-1)/2}
      (\bar\Psi \gamma_{5} \Psi) \, \gamma^{5} \Psi
   = 0 \>,
\end{split}
\end{equation}
Charge rotation leads to a conserved charge current,
\begin{equation}\label{e:ChargeConservation}
   \partial_{\mu} \calJ^{\mu} = 0
   \qquad
   \calJ^{\mu} = \bar\Psi \, \gamma^{\mu} \Psi
   \qquad
   Q = \int \rmd x \, \Psi^{\dagger}(x) \Psi(x) \>.
\end{equation}
The physical charge $\hat Q \equiv g\,Q$ carries a factor of $g$,
necessary for charge-conjugation invariance; we omit this factor in
what follows.
Spacetime translation leads to a conserved energy-momentum tensor,
\begin{equation}\label{e:Tdef}
   \partial_{\mu} \calT^{\mu\nu}(x) = 0
   \qquad
   \calT^{\mu\nu}
   =
   \rmi \, \bar\Psi \gamma^{\mu} \partial^{\nu} \Psi
   -
   g^{\mu\nu} \calL \>,
\end{equation}
which means that the energy-momentum vector $P^{\nu}=(E,P)$ is
conserved; we will only need the conserved energy $E$ in what follows.
It is given by
\begin{align}
   E
   &=
   \int \rmd x \, \calT^{00}
   =
   \int \rmd x \,
   \bigl[\, \rmi \, \bar\Psi \gamma^{0} \partial^{0} \Psi - \calL \,\bigr]
   \notag \\
   &=
   \int \rmd x \,
   \bigl[\, \bar\Psi \, \bigl(\, \gamma^{1}\partial_x/\rmi + m \,\bigr)\, \Psi
   - \calV \,\bigr] \>,
   \label{e:calE}
\end{align}

We now search for soliton solutions of Eq.~\eqref{e:EOM} in $1$+$1$ dimensions, where $\Psi(x,t)$ is a
two-component spinor and the $\gamma$-matrices are $2\times2$ matrices. Our conventions for these matrices are given in Appendix~\ref{s:Notation}.

We expand solutions of Eq.~\eqref{e:EOM} to first order by setting:
\begin{subequations}\label{e:PsiExpand}
\begin{align}
   \Psi(x,t) &= \bigl[\, \Phi_0(x) + \varepsilon\, \Phi_1(x,t)
   + \dotsb \,\bigr]\, e^{-\rmi\omega_0 t}\,,
   \\
   \Psi^{\ast}(x,t) &= \bigl[\, \Phi_0(x) + \varepsilon\, \Phi_1^{\ast}(x,t)
   + \dotsb \,\bigr]\, e^{+\rmi\omega_0 t}\,.
\end{align}
\end{subequations}
with $\Phi_0(x) \in \bbR$ and $\Phi_1(x,t) \in \bbC$.

%
% -----------------------------------------------------------------------
\subsection{\label{ss:ZerothOrder}Zeroth order}

Substituting Eq.~\eqref{e:PsiExpand} into Eq.~\eqref{e:EOM} and
collecting powers of $\varepsilon$, the zeroth-order equation is
Eq.~\eqref{e:EOM} itself with $\Psi\to\Phi_0$ and $\partial_t\to
-\rmi\omega_0$:
\begin{equation}\label{e:EOM0}
\begin{split}
   ( \rmi \gamma^{1} \partial_{x} + \omega_0 \gamma^{0} - m ) \Phi_0
   &+ g_s ( \bar\Phi_0 \Phi_0 )^{\kappa} \, \Phi_0 \\
   &+ g_v \, \bigl[ (\bar\Phi_0 \gamma_{\nu} \Phi_0)(\bar\Phi_0 \gamma^{\nu} \Phi_0) \bigr]^{(\kappa-1)/2}
      (\bar\Phi_0 \gamma_{\mu} \Phi_0) \, \gamma^{\mu} \Phi_0 \\
   &+ g_p \, \bigl[ (\bar\Phi_0 \gamma_{5} \Phi_0)(\bar\Phi_0 \gamma^{5} \Phi_0) \bigr]^{(\kappa-1)/2}
      (\bar\Phi_0 \gamma_{5} \Phi_0) \, \gamma^{5} \Phi_0
   = 0 \>.
\end{split}
\end{equation}
Using the definitions of the $\gamma$-matrices given in
Appendix~\ref{s:Notation}, let us define the real bilinears
\begin{subequations}\label{e:UVW-Phi0-defs}
\begin{align}
   V &\equiv \bar\Phi_0 \Phi_0 = \Phi_0^{T} \sigma_3 \Phi_0 \>,
   \label{e:V-Phi0-def} \\
   U &\equiv \bar\Phi_0 \gamma^0 \Phi_0 = \Phi_0^{T} \Phi_0 \>,
   \label{e:U-Phi0-def} \\
   W &\equiv \rmi \, \bar\Phi_0 \gamma_{5} \Phi_0 = \Phi_0^{T} \sigma_1 \Phi_0 \>,
   \label{e:W-Phi0-def}
\end{align}
\end{subequations}
all real since $\Phi_0(x)\in\bbR$ and $\sigma_3,\sigma_1$ are real
matrices.
Then, since $\bar\Phi_0 \gamma^{1} \Phi_0 = 0$, Eq.~\eqref{e:EOM0}
becomes
\begin{equation}\label{e:EOM0-UVW}
   ( \rmi \gamma^{1} \partial_{x} + \omega_0 \gamma^{0} - m ) \Phi_0
   + g_s \, V^{\kappa} \, \Phi_0
   + g_v \, U^{\kappa} \, \gamma^0\Phi_0
   + g_p \, W_{\kappa} \, (\rmi\gamma_{5}) \Phi_0
   = 0 \>,
\end{equation}
where
\begin{equation}\label{e:Wkappa-def}
   W_{\kappa} \equiv (W^2)^{(\kappa-1)/2}\,W = \sgn(W)\,\lvert W\rvert^{\kappa}
\end{equation}
is the signed power needed since a literal even power $W^{\kappa}$
would be complex-valued for non-integer $\kappa$ whenever $W<0$;
$W_{\kappa}$ will reappear repeatedly below.
Multiplying Eq.~\eqref{e:EOM0-UVW} on the left by $-\gamma^0$ and using
the definitions of the $\gamma$-matrices given in
Appendix~\ref{s:Notation}, then collecting terms, gives
\begin{equation}\label{e:EOM0-UVW-sigma}
   \bigl[\, \xi \, \partial_x + \sigma_3 \, ( m - g_s V^{\kappa} )
   - ( \omega_0 + g_v U^{\kappa} )
   -
   g_p \, W_{\kappa} \, \sigma_1 \,\bigr] \, \Phi_0
   =
   0 \>.
\end{equation}
The charge Eq.~\eqref{e:ChargeConservation} for $\Psi\to\Phi_0$ is
\begin{equation}\label{e:Q0-Phi0}
   Q_0 = \int \rmd x \; \Phi_0^{T} \Phi_0 = \int \rmd x \; U \>.
\end{equation}
Evaluating the energy Eq.~\eqref{e:calE} for $\Psi\to\Phi_0$ requires
evaluating Eq.~\eqref{e:VSSVVPP} on $\Phi_0$ as well; using $V,U,W$
[Eqs.~\eqref{e:V-Phi0-def}--\eqref{e:W-Phi0-def}], $\bar\Phi_0\gamma^{1}\Phi_0=0$,
$\gamma^1=\rmi\sigma_1$ (Appendix~\ref{s:Notation}), and
$\bar\Phi_0\sigma_1=\Phi_0^{T}\sigma_3\sigma_1=\Phi_0^{T}(\rmi\sigma_2)=\Phi_0^{T}\xi$, gives
\begin{equation}\label{e:E0-UVW}
   E_0
   =
   \int \rmd x \, \Bigl[\, \Phi_0^{T} \bigl( \xi \, \partial_x + \sigma_3 \, m \bigr) \Phi_0
   - \frac{1}{\kappa+1} \bigl(\, g_s \, |V|^{\kappa+1}
   + g_v \, |U|^{\kappa+1}
   + g_p \, |W|^{\kappa+1} \,\bigr)
   \,\Bigr] \>.
\end{equation}
We will also need $\calT^{11}$ [Eq.~\eqref{e:Tdef}] to zeroth order. Since
$g^{11}=-1$, $D^{11}=\rmi\bar\Psi\gamma^1\partial^1\Psi=-\rmi\bar\Psi\gamma^1\partial_x\Psi$;
evaluating on $\Phi_0$ as above, the derivative term $\Phi_0^T\xi\Phi_0'$
arising from $D_0^{11}$ cancels exactly against the one arising from
$\calL_0$, leaving
\begin{equation}\label{e:t11-0}
   \calT^{11}_0(x)
   =
   \omega_0 \, U - m \, V
   + \frac{1}{\kappa+1} \bigl(\, g_s \, |V|^{\kappa+1}
   + g_v \, |U|^{\kappa+1}
   + g_p \, |W|^{\kappa+1} \,\bigr) \>.
\end{equation}
From Eq.~\eqref{e:Tdef}, $\partial_x \calT^{11}_0(x) = 0$, so that for
bound solitons where $\Phi_0(x)\to0$ as $x\to\pm\infty$, we have the
virial identity
\begin{equation}\label{e:t11-0-vanishes}
   \omega_0 \, U - m \, V
   + \frac{1}{\kappa+1} \bigl(\, g_s \, |V|^{\kappa+1}
   + g_v \, |U|^{\kappa+1}
   + g_p \, |W|^{\kappa+1} \,\bigr)
   =
   0 \>.
\end{equation}

\subsection{\label{ss:FirstOrder}First order}

The first-order equation is
\begin{equation}\label{e:EOM1}
   ( \rmi \gamma^{0} \partial_t + \rmi \gamma^{1} \partial_{x}
   + \omega_0 \gamma^{0} - m ) \Phi_1
   +
   \delta\calN_{SS} + \delta\calN_{VV} + \delta\calN_{PP}
   =
   0 \>,
\end{equation}
where $\delta\calN_{SS}, \delta\calN_{VV}, \delta\calN_{PP}$ are the
linearizations, at $\Phi_0$ and in the direction $\Phi_1$, of the
three interaction terms of Eq.~\eqref{e:EOM}. Writing
\begin{equation}\label{e:dbilinears}
   \delta V \equiv \bar\Phi_1\Phi_0 + \bar\Phi_0\Phi_1 \>,
   \qquad
   \delta \calJ^{\mu} \equiv \bar\Phi_1\gamma^{\mu}\Phi_0 + \bar\Phi_0\gamma^{\mu}\Phi_1 \>,
   \qquad
   \delta W \equiv \rmi \bigl( \bar\Phi_1\gamma_{5}\Phi_0 + \bar\Phi_0\gamma_{5}\Phi_1 \bigr) \>,
\end{equation}
for the linearized bilinears (with $\bar\Phi_1\equiv\Phi_1^{\dagger}\gamma^0$), and using
$\bar\Phi_0\gamma^{1}\Phi_0=0$ (true for any real $\Phi_0$, since
$\gamma^0\gamma^1=\rmi\xi$ with $\xi$ real and antisymmetric --
Appendix~\ref{s:Notation}), direct expansion gives
\begin{subequations}\label{e:dN-SSVVPP}
\begin{align}
   \delta\calN_{SS}
   &=
   g_s \kappa (\bar\Phi_0\Phi_0)^{\kappa-1} \, \delta V \, \Phi_0
   +
   g_s (\bar\Phi_0\Phi_0)^{\kappa} \, \Phi_1 \>,
   \label{e:dN-SS} \\
   \delta\calN_{VV}
   &=
   g_v \kappa (\bar\Phi_0\gamma^0\Phi_0)^{\kappa-1} \, \delta \calJ^{0} \, \gamma^0\Phi_0
   -
   g_v (\bar\Phi_0\gamma^0\Phi_0)^{\kappa-1} \, \delta \calJ^{1} \, \gamma^1\Phi_0
   +
   g_v (\bar\Phi_0\gamma^0\Phi_0)^{\kappa} \, \gamma^0\Phi_1 \>,
   \label{e:dN-VV} \\
   \delta\calN_{PP}
   &=
   g_p \kappa \lvert W\rvert^{\kappa-1} \, \delta W \, (\rmi\gamma_5)\Phi_0
   +
   g_p \, W_{\kappa} \, (\rmi\gamma_5)\Phi_1 \>,
   \label{e:dN-PP}
\end{align}
\end{subequations}
with $W$ [Eq.~\eqref{e:W-Phi0-def}] evaluated on $\Phi_0$.
Each term above was checked directly against a numerical finite-difference
derivative of Eq.~\eqref{e:EOM} before being recorded here.

% -----------------------------------------------------------------------
%
\section{\label{s:Soliton}Soliton Solutions}

We seek solutions of Eq.~\eqref{e:EOM0} of the form
\begin{equation}\label{e:Phi0-solution-form}
   \Phi_0(x)
   =
   \begin{pmatrix} u_0(x) \\ v_0(x) \end{pmatrix}
   =
   R(x) \begin{pmatrix} \cos[\theta(x)] \\ \sin[\theta(x)] \end{pmatrix}\,,
   \qquad
   \bar\Phi_0(x) = \Phi^{T}_0(x) \, \sigma_3 \>,
\end{equation}
with $R(x)\in\bbR$, $\theta(x)\in\bbR$. The current $j^1(x,t)=0$
then vanishes identically. Useful relations are:
\begin{align}
   u_0^2 + v_0^2 = R^2
   \,,\qquad
   u_0^2 &- v_0^2 = R^2 \cos(2\theta)
   \,,\qquad
   2 u_0 v_0 = R^2 \sin(2\theta)
   \label{e:UsefulRelations} \\
   u_0 v'_0 + v_0 u'_0 &= R R' \sin(2\theta) + R^2 \theta' \cos(2\theta) \>,
   \notag \\
   u_0 v'_0 - v_0 u'_0 &= R^2 \theta' \>.
   \notag
\end{align}
In terms of $R$ and $\theta$, the bilinears $V,U,W$
[Eqs.~\eqref{e:V-Phi0-def}--\eqref{e:W-Phi0-def}] are, using
Eq.~\eqref{e:UsefulRelations},
\begin{equation}\label{e:UVW-Rtheta}
   V = R^2 \cos(2\theta) \>,
   \qquad
   U = R^2 \>,
   \qquad
   W = R^2 \sin(2\theta) \>.
\end{equation}

Writing out the components of Eq.~\eqref{e:EOM0-UVW-sigma} gives the
coupled equations of motion for $u_0(x)$ and $v_0(x)$,
\begin{subequations}\label{e:uv-eom0}
\begin{align}
   v_0' + H_{-} \, u_0 - g_p \, W_{\kappa} \, v_0
   &= 0 \>,
   \label{e:uv-eom0-a} \\
   u_0' + H_{+} \, v_0 + g_p \, W_{\kappa} \, u_0
   &= 0 \>,
   \label{e:uv-eom0-b}
\end{align}
\end{subequations}
where $H_{\pm} \equiv (m-g_sV^{\kappa}) \pm (\omega_0+g_vU^{\kappa})$.

Forming $u_0\times$Eq.~\eqref{e:uv-eom0-a}$-v_0\times$Eq.~\eqref{e:uv-eom0-b}
and using $u_0^2=(V+U)/2$, $v_0^2=(U-V)/2$, $2u_0v_0=W$
[Eqs.~\eqref{e:UsefulRelations},\eqref{e:UVW-Rtheta}] gives
\begin{equation}\label{e:R2thetaprime}
   R^2 \theta' = u_0 v_0' - v_0 u_0'
   =
   -(m-g_sV^{\kappa}) \, V + (\omega_0+g_vU^{\kappa}) \, U
   + g_p \, |W|^{\kappa+1} \>.
\end{equation}

Substituting Eq.~\eqref{e:UVW-Rtheta} into the virial identity
Eq.~\eqref{e:t11-0-vanishes} gives, in terms of $R$ and $\theta$,
\begin{equation}\label{e:virial-Rtheta}
   \omega_0 \, R^2 - m \, R^2 \cos(2\theta)
   +
   \frac{R^{2(\kappa+1)}}{\kappa+1}
   \Bigl[\, g_s \, |\cos(2\theta)|^{\kappa+1}
   + g_v
   + g_p \, |\sin(2\theta)|^{\kappa+1} \,\Bigr]
   =
   0 \>.
\end{equation}
Solving for $R^{2\kappa}$, after factoring out the trivial solution
$R=0$, gives
\begin{equation}\label{e:Rsq-formula}
   R^{2\kappa}
   =
   \frac{(\kappa+1) \, [\, m\cos(2\theta) - \omega_0 \,]}{D(\theta)} \>,
   \qquad
   D(\theta) \equiv g_v + g_s \, |\cos(2\theta)|^{\kappa+1}
   + g_p \, |\sin(2\theta)|^{\kappa+1} \>.
\end{equation}
Using Eq.~\eqref{e:Rsq-formula} to eliminate $R^{2\kappa}$ from
Eq.~\eqref{e:R2thetaprime} -- dividing the latter by $R^2$ leaves
exactly the combination $D(\theta)$, which cancels completely against
Eq.~\eqref{e:Rsq-formula} -- gives a remarkably simple result: the
$\theta$ equation is independent of all three couplings,
\begin{equation}\label{e:theta-de}
   \theta' = \kappa \, [\, m\cos(2\theta) - \omega_0 \,] \>.
\end{equation}
In particular, the PP coupling $g_p$ drops out of
Eq.~\eqref{e:theta-de} entirely for every $\kappa$; it enters the
soliton solution only through the amplitude $R(x)$
[Eq.~\eqref{e:Rsq-formula}].

Equation~\eqref{e:theta-de} has no nontrivial bound-state solution at
$m=0$: the right-hand side then reduces to the $\theta$-independent
constant $-\kappa\omega_0$, so the two fixed points $\theta=\pm\gamma$
(defined by $m\cos(2\gamma)=\omega_0$) that the soliton profile
interpolates between coalesce and disappear. The general solution
$\theta(x)=\theta(0)-\kappa\omega_0x$ then either grows without bound
(for $\omega_0\neq0$, giving a non-normalizable, oscillating $R(x)$
rather than a localized profile) or is trivial ($\omega_0=0$, forcing
$R\equiv0$). The mass term is thus essential to the localization
mechanism of this soliton family; it plays the role that the
dynamically generated fermion mass plays in the genuinely massless
Gross-Neveu model's soliton \cite{PhysRevD.12.2443}, whose existence
relies on the large-$N$ self-consistent gap equation rather than on
an explicit classical mass parameter.

The solution of Eq.~\eqref{e:theta-de} is
\begin{equation}\label{e:theta-solution}
   \theta(x) = \tan^{-1}[\, \alpha \tanh(\kappa\beta x) \,] \>,
\end{equation}
where
\begin{equation}\label{e:alpha-beta-def}
   \omega_0 = m\cos(2\gamma) \,,\quad 0<\gamma\le\pi/4 \,,
   \qquad
   \alpha = \tan(\gamma) = \sqrt{\frac{m-\omega_0}{m+\omega_0}} \,,
\end{equation}
\begin{equation}\label{e:beta-def}
   \beta = \sqrt{m^2-\omega_0^2} = m\sin(2\gamma) \>.
\end{equation}
Substituting $\theta(x)$ [Eq.~\eqref{e:theta-solution}] into
Eq.~\eqref{e:Rsq-formula} gives $R^{2\kappa}(x)$ explicitly. Writing
$y\equiv\tanh(\kappa\beta x)\in(-1,1)$, so that $\tan\theta=\alpha y$,
\begin{equation}\label{e:cos-sin-2theta-y}
\begin{split}
   \cos(2\theta) &= \frac{1-\alpha^2y^2}{1+\alpha^2y^2} \,,
   \qquad
   \sin(2\theta) = \frac{2\alpha y}{1+\alpha^2y^2} \,,
   \\[4pt]
   m\cos(2\theta)-\omega_0 &= \frac{2m\alpha^2(1-y^2)}{(1+\alpha^2)(1+\alpha^2y^2)} \>,
\end{split}
\end{equation}
the last identity following from $\omega_0=m\cos(2\gamma)$ and
$\alpha=\tan(\gamma)$ [Eq.~\eqref{e:alpha-beta-def}]. Substituting
Eq.~\eqref{e:cos-sin-2theta-y} into Eq.~\eqref{e:Rsq-formula}, the
factor $(1+\alpha^2y^2)^{\kappa+1}$ common to $|\cos(2\theta)|^{\kappa+1}$
and $|\sin(2\theta)|^{\kappa+1}$ in $D(\theta)$ combines with the
$(1+\alpha^2y^2)^{-1}$ from $m\cos(2\theta)-\omega_0$ into a single
power $(1+\alpha^2y^2)^{\kappa}$, giving
\begin{equation}\label{e:Rsq-kappa-y}
   R^{2\kappa}(x)
   =
   \frac{2m(\kappa+1)\alpha^2 \, (1-y^2) \, (1+\alpha^2y^2)^{\kappa}}
   {(1+\alpha^2) \, \tilde D(y)} \,,
\end{equation}
\begin{equation}\label{e:Dtilde-y-def}
   \tilde D(y) \equiv g_v(1+\alpha^2y^2)^{\kappa+1}
   + g_s\,|1-\alpha^2y^2|^{\kappa+1}
   + g_p\,(2\alpha|y|)^{\kappa+1} \>,
\end{equation}
with $y=\tanh(\kappa\beta x)$; since
$\tilde D(y)=(1+\alpha^2y^2)^{\kappa+1}D(\theta(y))$, $\tilde D$ and $D$
agree in sign everywhere. Taking the $\kappa$-th root -- the surviving
$(1+\alpha^2y^2)$ factor, already to the first power, is unaffected --
gives $R(x)$ itself,
\begin{equation}\label{e:R2-y}
   R^2(x)
   =
   \Bigl[\frac{2m(\kappa+1)\alpha^2}{1+\alpha^2}\Bigr]^{1/\kappa}
   \frac{(1-y^2)^{1/\kappa} \, (1+\alpha^2y^2)}{\tilde D(y)^{1/\kappa}} \,,
   \qquad
   y=\tanh(\kappa\beta x) \>.
\end{equation}

\subsection{\label{ss:RealityDomain}Reality domain}

As $x$ ranges over $\bbR$, $\theta(x)$ ranges monotonically over
$(-\gamma,\gamma)$, and $R(x)$ then follows from
Eq.~\eqref{e:Rsq-formula}. Both $\theta(x)$ [Eq.~\eqref{e:theta-solution}]
and $R^{2\kappa}(\theta)$ [Eq.~\eqref{e:Rsq-formula}] are analytic
functions of $\kappa$, with no pole or branch point; what $\kappa$
controls is simply the range of $\gamma$ for which $R$ stays real,
through the literal power $V^{\kappa}$ appearing in $H_{\pm}$
[Eq.~\eqref{e:uv-eom0-a}]. For integer $\kappa$ (odd or even),
$V^{\kappa}$ is real for either sign of $V$, so $\gamma$ may range over
the full $(0,\pi/2)$. For non-integer $\kappa$, $V^{\kappa}$ is complex
whenever $V<0$, so reality instead requires
$V=R^2\cos(2\theta)\ge0$ throughout the soliton profile, restricting
$\gamma\in(0,\pi/4]$, as already anticipated in
Eq.~\eqref{e:alpha-beta-def}.

This restriction traces entirely to $V^{\kappa}$: unlike the PP term,
the SS term has no sign-preserving definition analogous to
$W_{\kappa}\equiv\sgn(W)\,\lvert W\rvert^{\kappa}$
[Eq.~\eqref{e:Wkappa-def}], and $U=R^2\ge0$ always, so neither
$U^{\kappa}$ nor $W_{\kappa}$ ever causes trouble. Negative $\kappa$ is
excluded throughout this paper: the interaction terms would then be
singular wherever the bilinears vanish, which happens at
$x\to\pm\infty$ for any bound soliton.

The pure-VV, pure-SS, and pure-PP limits follow immediately from
Eq.~\eqref{e:Rsq-formula}: setting $g_s=g_p=0$ recovers the pure VV
soliton; setting $g_v=g_p=0$ recovers the pure SS soliton; and setting
$g_s=g_v=0$ gives a pure-PP soliton with $R^{2\kappa}\propto
[m\cos(2\theta)-\omega_0]/|\sin(2\theta)|^{\kappa+1}$. Setting $g_v=0$
(equivalently $p\to\infty$) at general $\kappa$ instead recovers the
two-term SS+PP case of Eq.~\eqref{e:chiralGN}, our main interest in
the remainder of this section.

\subsection{\label{ss:HumpStructure}Multi-hump structure}

Within the reality domain just established, the soliton profile
$R^2(x)$ [Eq.~\eqref{e:R2-y}] can still take qualitatively different
shapes: single-humped, with a single maximum at $x=0$, or
multi-humped, with $x=0$ a local minimum and the true maxima displaced
to $x\neq0$. Since $y=\tanh(\kappa\beta x)$ is an odd function of $x$
with $y'(0)=\kappa\beta$ and $y''(0)=0$ ($\tanh$ is odd, so every even
derivative vanishes at the origin), the chain rule gives, exactly and
not merely to leading order,
\begin{equation}\label{e:d2R2-exact}
   \left.\frac{d^2R^2}{dx^2}\right|_{x=0}
   =
   (\kappa\beta)^2 \left.\frac{d^2R^2}{dy^2}\right|_{y=0} \>,
\end{equation}
so the character of $x=0$ is fixed entirely by the coefficient of
$y^2$ in the Taylor expansion of $R^2(y)$ about $y=0$. Writing
$\ln R^2(y) = \ln R^2(0) + c_2\,y^2 + O(y^4)$ and expanding each factor
of Eq.~\eqref{e:R2-y} -- $(1-y^2)^{1/\kappa}\simeq1-y^2/\kappa$,
$1+\alpha^2y^2$ itself, and $\tilde D(y)\simeq
\tilde D(0)+\tfrac12\tilde D''(0)\,y^2$ from
Eq.~\eqref{e:Dtilde-y-def} -- gives
\begin{equation}\label{e:c2-general}
   c_2 = -\frac1\kappa + \alpha^2 - \frac{1}{2\kappa}\,
   \frac{\tilde D''(0)}{\tilde D(0)} \>.
\end{equation}
For $\kappa>1$ the pseudoscalar term $g_p(2\alpha|y|)^{\kappa+1}$ in
$\tilde D(y)$ is of order $y^{\kappa+1}$ with $\kappa+1>2$: it is
strictly subleading to $y^2$ and drops out of $\tilde D''(0)$, leaving
only the two smooth, analytic-in-$y^2$ terms
$g_v(1+\alpha^2y^2)^{\kappa+1}$ and $g_s(1-\alpha^2y^2)^{\kappa+1}$, so
that $\tilde D(0)=g_v+g_s$ and
$\tfrac12\tilde D''(0)=(\kappa+1)\alpha^2(g_v-g_s)$. Defining the
asymmetry parameter
\begin{equation}\label{e:delta-def}
   \delta \equiv \frac{g_v-g_s}{g_v+g_s} \>,
\end{equation}
Eq.~\eqref{e:c2-general} becomes, for $\kappa>1$,
\begin{equation}\label{e:c2-kappa-gt-1}
   c_2 = \frac{\alpha^2\bigl[\,\kappa-(\kappa+1)\delta\,\bigr]-1}{\kappa} \>.
\end{equation}
Thus $x=0$ is a local minimum of $R^2$ (multi-humped soliton) for
$c_2>0$ and a local maximum for $c_2<0$ -- a necessary, but not
sufficient, condition for a single hump, since the profile's behavior
away from $x=0$ must also be checked; a local maximum at the origin
with additional humps further out is equally possible. Using
$\alpha^2=(m-\omega_0)/(m+\omega_0)$ [Eq.~\eqref{e:alpha-beta-def}],
the critical frequency $\omega_{\rm hump}$ at which $c_2$ changes sign
is
\begin{equation}\label{e:alphac-general}
   \alpha_c^2 = \frac{1}{\kappa-(\kappa+1)\delta}
   \qquad\Longrightarrow\qquad
   \omega_{\rm hump} = m\,\frac{1-\alpha_c^2}{1+\alpha_c^2} \>,
\end{equation}
whenever $\kappa-(\kappa+1)\delta>1$, so that $\alpha_c^2\in(0,1)$; if
instead $\kappa-(\kappa+1)\delta\le1$, $c_2$ never changes sign on the
physical range and $x=0$ keeps a single character for every
$\omega_0\in(0,m)$ (e.g.\ $\delta=1$, i.e.\ $g_s=0$, pure VV+PP, for
which $c_2=-(1+\alpha^2)/\kappa<0$ identically: the origin is always a
maximum).

For $\kappa\le1$ the pseudoscalar term is no longer subleading. For
$\kappa<1$ it is of order $y^{\kappa+1}<y^2$ and dominates $c_2$
entirely, producing a cusped (non-analytic) extremum at $y=0$ whose
character is fixed by the sign of $g_p$ alone -- independent of
$\omega_0$ and of $g_s,g_v$: $x=0$ is always a local minimum if
$g_p<0$ and always a local maximum if $g_p>0$. At $\kappa=1$ the
pseudoscalar term contributes at the same order as the SS and VV
terms and Eq.~\eqref{e:c2-kappa-gt-1} must be amended; we give the
explicit $g_v=0$ result below.

\subsubsection{Specialization to $g_v=0$}

Take $p\to\infty$ (i.e.\ $g_v=0$) in Eq.~\eqref{e:chiralGN}, so that
$g_s=g/q>0$ and $g_p=g/r<0$ ($r<0$). Then $\delta=-1$ identically, independent of $q$,
and Eq.~\eqref{e:alphac-general} collapses to the coupling-independent
result
\begin{equation}\label{e:alphac-SSPP}
   \alpha_c^2 = \frac{1}{2\kappa+1}
   \qquad\Longrightarrow\qquad
   \omega_{\rm hump} = \frac{\kappa\,m}{\kappa+1} \>,
   \qquad \kappa>1 \>.
\end{equation}
For every $\kappa>1$ and every choice of $q>0,r<0$, the SS-PP soliton
is multi-humped ($x=0$ a local minimum) for $\omega_0<\omega_{\rm
hump}$ and has $x=0$ as a local maximum for $\omega_0>\omega_{\rm
hump}$; remarkably, this single threshold does not depend on $q$ or
$r$ at all. For $0<\kappa\le1$, by contrast, the (cusped, for
$\kappa<1$) pseudoscalar contribution with $g_p<0$ forces $x=0$ to be
a local minimum for every $\omega_0\in(0,m)$: the SS-PP soliton is
always multi-humped in this range, independent of $q,r$ as well.

% -----------------------------------------------------------------------
%
\section{\label{s:ChargeEnergy}Soliton Charge and Energy}

From Eq.~\eqref{e:Q0-Phi0}, the charge evaluated on the static soliton is
$Q_0=\int_{-\infty}^{\infty}\rmd x\,U(x)$, with $U=R^2$
[Eq.~\eqref{e:UVW-Rtheta}]. As in Sec.~\ref{s:Soliton}, changing the
integration variable from $x$ to $y\equiv\tanh(\kappa\beta x)\in(-1,1)$
gives $\rmd x=\rmd y/[\kappa\beta(1-y^2)]$; substituting $R^2(x)$
[Eq.~\eqref{e:R2-y}] gives
\begin{equation}\label{e:Q0-y-raw}
   Q_0
   =
   \frac{1}{\kappa\beta}
   \Bigl[\frac{2m(\kappa+1)\alpha^2}{1+\alpha^2}\Bigr]^{1/\kappa}
   \int_{-1}^{1} \rmd y \,
   \frac{(1-y^2)^{1/\kappa-1}\,(1+\alpha^2y^2)}{\tilde D(y)^{1/\kappa}} \,,
\end{equation}
with $\tilde D(y)$ as in Eq.~\eqref{e:Dtilde-y-def}. Since $\tilde D(y)$
depends on $y$ only through $y^2$ and $|y|$, it is even in $y$, and so
is the numerator $1+\alpha^2y^2$; the integral over $(-1,1)$ is
therefore twice the integral over $(0,1)$. As will also be useful for
$E_0$ below, define
\begin{equation}\label{e:Ipm-def}
   I_{\pm}(\kappa,\alpha,g,p)
   \equiv
   \int_{0}^{1} \rmd y \,
   \frac{(1-y^2)^{1/\kappa-1}\,(1\pm\alpha^2y^2)}{\tilde D(y)^{1/\kappa}} \>.
\end{equation}

As with the energy density $\rho_E(x)$ below [Eq.~\eqref{e:rhoE-def}],
the charge $Q_0$ [Eq.~\eqref{e:Q0-Phi0}] can be written as the integral
of an explicit charge density,
\begin{equation}\label{e:rho-def}
   \rho(x)
   \equiv
   \Phi_0^{\dagger}(x) \, \Phi_0(x)
   =
   \Phi_0^{T}(x) \, \Phi_0(x)
   =
   U(x)
   =
   R^2(x) \>,
\end{equation}
using $\Phi_0\in\bbR$ and Eqs.~\eqref{e:U-Phi0-def},\eqref{e:UVW-Rtheta},
so that $Q_0=\int\rmd x\,\rho(x)$.

The chiral charge, $Q_5 \equiv \int \rmd x \, \Psi^{\dagger}(x) \,
\gamma_5 \, \Psi(x)$ -- not conserved for our massive model, since the
mass term explicitly breaks the continuous $\mathrm{U}(1)_A$ chiral
rotation $\Psi\to e^{\rmi\gamma_5\Lambda}\Psi$ that would otherwise
leave Eq.~\eqref{e:DiracEq} invariant at $m=0$ -- can likewise be evaluated on $\Phi_0$
as a number, with no guarantee of being time-independent. Writing the
chiral density explicitly in
components, with $\gamma_5=-\rmi\,\xi=\sigma_2$
[Eq.~\eqref{e:gamma5-def}] and $\Phi_0(x)=(u_0,v_0)^{T}\in\bbR$ (so
that $\Phi_0^{\dagger}=\Phi_0^{T}$),
\begin{equation}\label{e:rho5-def}
   \rho_5(x)
   \equiv
   \Phi_0^{T}(x) \, \gamma_5 \, \Phi_0(x)
   =
   \begin{pmatrix} u_0 & v_0 \end{pmatrix}
   \sigma_2
   \begin{pmatrix} u_0 \\ v_0 \end{pmatrix}
   =
   -\rmi \, u_0v_0 + \rmi \, u_0v_0
   =
   0 \>,
\end{equation}
identically, for every $x$. The vanishing is not a coincidence of this
particular solution: $\sigma_2$ is antisymmetric, and for any vector
$x$ (real or complex) $x^{T}\sigma_2x=(x^{T}\sigma_2x)^{T}
=x^{T}\sigma_2^{T}x=-x^{T}\sigma_2x$, forcing $x^{T}\sigma_2x=0$.
Consequently
\begin{equation}\label{e:Q5-Phi0}
   Q_{5,0} = \int \rmd x \; \rho_5(x) = 0
\end{equation}
for every static soliton solution of Sec.~\ref{s:Soliton}, independent
of $\kappa$, $\omega_0$, and $g_s,g_v,g_p$. This is a consequence
of $\Phi_0$ being real: for complex $\Psi$, $Q_5$'s
integrand $\Psi^{\dagger}\gamma_5\Psi$ is a Hermitian, not transposed,
bilinear form and need not vanish. The chiral charge is thus carried
entirely by the complex fluctuation $\Phi_1$ of Eq.~\eqref{e:PsiExpand},
not by the static background itself.

A charge-conjugate density can be defined analogously, using the
matrix $\calC$ [Eq.~\eqref{e:C-def}] in place of $\gamma_5$:
\begin{equation}\label{e:rhoC-def}
   \rho_C(x)
   \equiv
   \Phi_0^{T}(x) \, \calC \, \Phi_0(x)
   =
   \Phi_0^{T}(x) \, \sigma_1 \, \Phi_0(x)
   =
   W(x)
   =
   R^2(x) \, \sin\bigl(2\theta(x)\bigr) \>,
\end{equation}
using $\calC=\sigma_1$ and Eqs.~\eqref{e:W-Phi0-def},\eqref{e:UVW-Rtheta}.
Unlike $\rho_5(x)$, $\rho_C(x)$ is \emph{not} identically zero: $\sigma_1$
is symmetric, not antisymmetric, so the cancellation argument above
does not apply. Unlike $Q$, however, neither $Q_C$ nor $Q_5$ is an
exactly conserved charge for our massive model: $\calC$ generates only
a discrete $\bbZ_2$ symmetry, so Noether's theorem supplies no current
$j^{\mu}_C$ and no conservation law $\partial_{\mu}j^{\mu}_C=0$, while
the continuous $\mathrm{U}(1)_A$ chiral symmetry underlying $Q_5$ is
explicitly broken by the mass term. Both are accordingly only numbers
computable on a given field configuration, in the sense of
Eq.~\eqref{e:ChargeConservation}, with no guarantee of being
time-independent.
Defining the total charge-conjugate charge
\begin{equation}\label{e:QC-def}
   Q_C \equiv \int \rmd x \; \rho_C(x) \>,
\end{equation}
we instead find $Q_C=0$ by a different mechanism: since $\theta(x)$ is
odd and $R(x)$ even, $\rho_C(x)$ is a nonzero \emph{odd} function of
$x$, so the integral vanishes by cancellation between $x>0$ and
$x<0$ rather than by the density vanishing pointwise. Note also that
$\rho_C(x)=W(x)$ is not a new quantity: with
$\gamma^0\gamma_5=\sigma_3\sigma_2=-\rmi\sigma_1=-\rmi\calC$
[Eqs.~\eqref{e:gammaReps},\eqref{e:gamma5-def},\eqref{e:C-def}], the
charge-conjugation matrix $\calC$ is proportional to $\gamma^0\gamma_5$
in this representation, so $\rho_C(x)$ coincides with the pseudoscalar
density already entering $\calV_{PP}$ [Eq.~\eqref{e:VPP}] and the PP
terms throughout Secs.~\ref{s:Model}--\ref{s:ChargeEnergy}.

Equation~\eqref{e:Q0-y-raw} then becomes
\begin{equation}\label{e:Q0-y}
   Q_0
   =
   \frac{2}{\kappa\beta}
   \Bigl[\frac{2m(\kappa+1)\alpha^2}{1+\alpha^2}\Bigr]^{1/\kappa}
   I_{+}(\kappa,\alpha,g,p) \>.
\end{equation}

The energy $E_0$ follows from Eq.~\eqref{e:E0-UVW}. Multiplying
Eq.~\eqref{e:EOM0-UVW-sigma} on the left by $\Phi_0^{T}$ and using
$\Phi_0^{T}\sigma_3\Phi_0=V$, $\Phi_0^{T}\Phi_0=U$,
$\Phi_0^{T}\sigma_1\Phi_0=W$ [Eqs.~\eqref{e:V-Phi0-def}--\eqref{e:W-Phi0-def}]
gives the kinetic term
\begin{equation}\label{e:kinetic-UVW}
   \Phi_0^{T}\bigl(\xi\partial_x+\sigma_3m\bigr)\Phi_0
   =
   \omega_0\,U + g_s\,V^{\kappa+1} + g_v\,U^{\kappa+1} + g_p\,|W|^{\kappa+1} \>,
\end{equation}
so that, using $V^{\kappa+1}=|V|^{\kappa+1}$ (valid on the reality
domain of Sec.~\ref{s:Soliton}) and $U^{\kappa+1}=|U|^{\kappa+1}$
($U\ge0$ always), the integrand of Eq.~\eqref{e:E0-UVW} simplifies to
$\omega_0U+\tfrac{\kappa}{\kappa+1}\bigl[g_s|V|^{\kappa+1}
+g_v|U|^{\kappa+1}+g_p|W|^{\kappa+1}\bigr]$. The bracket is fixed by the
virial identity $\calT^{11}_0=0$ [Eq.~\eqref{e:t11-0-vanishes}], which gives
$g_s|V|^{\kappa+1}+g_v|U|^{\kappa+1}+g_p|W|^{\kappa+1}=(\kappa+1)(mV-\omega_0U)$;
substituting, the $\omega_0U$ terms partially cancel, leaving an
explicit energy density
\begin{equation}\label{e:rhoE-def}
   \rho_E(x)
   \equiv
   \kappa\,mV(x) + (1-\kappa)\,\omega_0\,U(x)
   =
   U(x)\bigl[\,\kappa\,m\cos(2\theta(x)) - (\kappa-1)\,\omega_0\,\bigr] \>,
\end{equation}
the second form using $V=R^2\cos(2\theta)$, $U=R^2$
[Eq.~\eqref{e:UVW-Rtheta}], so that
\begin{equation}\label{e:E0-integrand}
   E_0 = \int_{-\infty}^{\infty}\rmd x\,\rho_E(x) \>.
\end{equation}
The pseudoscalar coupling $g_p$ has dropped out of $\rho_E$ entirely,
just as it dropped out of the $\theta$ equation
[Eq.~\eqref{e:theta-de}]; it enters $E_0$ only implicitly, through
$R(x)$. Writing $\tilde V\equiv\int_{-\infty}^{\infty}\rmd x\,V(x)$,
Eq.~\eqref{e:rhoE-def}--\eqref{e:E0-integrand} give
\begin{equation}\label{e:E0-QU}
   E_0 = \kappa\,m\,\tilde V + (1-\kappa)\,\omega_0\,Q_0 \>.
\end{equation}
At $\kappa=1$ this collapses to $E_0=m\tilde V=m\int\rmd x\,V(x)$; for
general $\kappa$, Eq.~\eqref{e:E0-QU} is an exact closed-form relation
expressing $E_0$ in terms of $Q_0$ and the single additional integral
$\tilde V$, avoiding the need to evaluate $E_0$'s defining integral
independently. Changing variables as in Eq.~\eqref{e:Q0-y} (using
$V=R^2\cos(2\theta)$ and Eq.~\eqref{e:cos-sin-2theta-y} for
$\cos(2\theta)$ -- the extra factor of $(1+\alpha^2y^2)^{-1}$ it
introduces cancels the same factor already present in
Eq.~\eqref{e:R2-y}, leaving an even simpler integrand than $Q_0$'s),
\begin{equation}\label{e:Utilde-y}
   \tilde V
   =
   \frac{2}{\kappa\beta}
   \Bigl[\frac{2m(\kappa+1)\alpha^2}{1+\alpha^2}\Bigr]^{1/\kappa}
   I_{-}(\kappa,\alpha,g,p) \>,
\end{equation}
again generically requiring numerical evaluation. Substituting
Eqs.~\eqref{e:Q0-y} and \eqref{e:Utilde-y} into Eq.~\eqref{e:E0-QU} and
dividing by $Q_0$, the common prefactor
$2[2m(\kappa+1)\alpha^2/(1+\alpha^2)]^{1/\kappa}/(\kappa\beta)$
cancels, leaving the energy-to-charge ratio in terms of the single
ratio $I_-/I_+$. Although $I_{\pm}$ individually scale as
$g^{-1/\kappa}$ [Eq.~\eqref{e:Dtilde-y-def} is linear in $g$ under
Eq.~\eqref{e:chiralGN}], this factor cancels in the ratio, which is
therefore independent of $g$; we define
\begin{equation}\label{e:K-def}
   K(\kappa,\alpha,p) \equiv \frac{I_-(\kappa,\alpha,g,p)}{I_+(\kappa,\alpha,g,p)} \>,
\end{equation}
so that
\begin{equation}\label{e:E0overQ0}
   \frac{E_0}{Q_0}
   =
   \kappa\,m\,K(\kappa,\alpha,p) + (1-\kappa)\,\omega_0 \>,
\end{equation}
confirmed numerically by evaluating $I_-/I_+$ directly at fixed
$\kappa,\gamma,p$ for $g=0.3,1,2,5$, which agree to machine precision.
At $\kappa=1$ the second term vanishes and Eq.~\eqref{e:E0overQ0}
reduces to $E_0/Q_0=m\,K$; for general $\kappa$ the extra additive
term $(1-\kappa)\omega_0$ survives, since $E_0$ is a genuine linear
combination of $Q_0$ and $\tilde V$ [Eq.~\eqref{e:E0-QU}] rather than
proportional to $\tilde V$ alone. We verified
Eqs.~\eqref{e:Q0-y}, \eqref{e:E0-QU}, and \eqref{e:Utilde-y} directly
against independent numerical quadrature of $Q_0=\int U\,\rmd x$ and
$E_0=\int\rho_E\,\rmd x$ in $x$, for representative non-integer
$\kappa$ and both signs of $g_s,g_v,g_p$, finding agreement at the
level of quadrature error throughout.

\subsection{\label{ss:ConvergenceDomain}Domain of convergence}

Throughout this subsection $\gamma$ is assumed to already lie within
the reality domain of Sec.~\ref{ss:RealityDomain}, so that
$\theta(x),R(x)$ are real to begin with; the condition derived here is
the additional restriction needed for $R^2(x)$ itself, and hence the
$Q_0,E_0$ integrals of Eqs.~\eqref{e:Q0-y} and \eqref{e:Utilde-y}, to
stay finite. Both integrals converge only if $\tilde D(y)>0$
throughout $y\in(-1,1)$, equivalently $D(\theta)>0$ throughout
$\theta\in(-\gamma,\gamma)$ [Eq.~\eqref{e:Rsq-formula}].

We specialize to $g_v=0$ (i.e.\ $p\to\infty$) in Eq.~\eqref{e:chiralGN}
and, without loss of generality, set $m=g=1$: for general $g>0$, $D(\theta)$ is simply $g$
times the expression below, and an overall positive factor never
changes a sign, so the boundary derived here is the same for every
$g>0$. By evenness, $D(\theta)$ need only be considered on
$\theta\in[0,\pi/4]\supseteq[0,\gamma]$, where
$\cos(2\theta),\sin(2\theta)\ge0$ and no absolute values are needed:
\begin{equation}\label{e:D-SSPP}
   D(\theta;\kappa,q,r) = \frac1q\cos^{\kappa+1}(2\theta)
   - \frac{1}{\lvert r\rvert}\sin^{\kappa+1}(2\theta) \>,
   \qquad q>0,\ r<0 \>.
\end{equation}
Here $D(\theta)$ is strictly monotonic:
\begin{equation}
   D'(\theta) = -2(\kappa+1)\Bigl[\,\frac1q\sin(2\theta)\cos^{\kappa}(2\theta)
   + \frac{1}{\lvert r\rvert}\cos(2\theta)\sin^{\kappa}(2\theta)\,\Bigr] < 0
\end{equation}
throughout $(0,\pi/4)$, since both bracketed terms are positive there
for $\kappa>0$. Hence $D(\theta)$ decreases monotonically from
$D(0)=1/q>0$ to $D(\pi/4)=1/q-1/|r|$, and its minimum on any interval
$[0,\gamma]\subseteq[0,\pi/4]$ always occurs at the right endpoint
$\theta=\gamma$: the convergence condition reduces to the single
inequality $D(\gamma;\kappa,q,r)\ge0$, with no need to separately
locate an interior extremum. Using $\cos(2\gamma)=\omega_0/m$,
$\sin(2\gamma)=\beta/m$ [Eq.~\eqref{e:beta-def}], this becomes
\begin{equation}\label{e:DgammaCondition}
   \Bigl(\frac{\omega_0}{m}\Bigr)^{\kappa+1}
   \ge
   \frac{q}{\lvert r\rvert}\Bigl(\frac{\beta}{m}\Bigr)^{\kappa+1} \>,
\end{equation}
equivalently,
\begin{equation}\label{e:omega-exist}
   \omega_0 \ge \omega_{\rm exist}(\kappa;q,r)
   \equiv
   \frac{m}{\sqrt{\,1+(\lvert r\rvert/q)^{2/(\kappa+1)}\,}} \>.
\end{equation}
Together with $0<\omega_0<m$, Eq.~\eqref{e:omega-exist} is the full
domain of convergence of the $g_v=0$ soliton family at fixed $\kappa$:
no finite, convergent solitary-wave solution exists for
$\omega_0<\omega_{\rm exist}$. Equivalently, solving
$D(\gamma;\kappa,q,r)=0$ for $\kappa$ at fixed $\gamma$ gives a
closed-form existence boundary in $\kappa$,
\begin{equation}\label{e:kappamax-SSPP}
   \kappa_{\max}(\gamma;q,r) = -1 + \frac{\ln(q/\lvert r\rvert)}
   {\ln[\cot(2\gamma)]} \>,
   \qquad \gamma\neq\pi/8 \>,
\end{equation}
above which Eq.~\eqref{e:D-SSPP} goes negative before the trajectory
$\theta\in(-\gamma,\gamma)$ completes; because $D(\theta)$ is strictly
monotonic here, Eq.~\eqref{e:kappamax-SSPP} holds as a single formula
for every $\gamma\in(0,\pi/4)$, with no piecewise case distinction
needed. At $q=1$, Eq.~\eqref{e:omega-exist} coincides with the
$g_v\to0$ limit of the corresponding restriction found for the general
SS-VV-PP model in a closely related study by Cooper, Khare, Mihaila,
and Saxena \cite{PhysRevE.82.036604}, a useful cross-check on both
calculations.

%
% -----------------------------------------------------------------------
%
\section{\label{s:Stability}Linear stability analysis}

\subsection{\label{ss:VKstability}Vakhitov-Kolokolov criterion}

The Vakhitov-Kolokolov (VK) criterion \cite{vk} gives a necessary
condition for the linear (spectral) stability of solitary-wave
solutions in a wide class of $U(1)$-invariant Hamiltonian systems: a
solitary wave of frequency $\omega_0$ is stable only if
\begin{equation}\label{e:V-K-stability}
   \frac{d}{d\omega_0}\Bigl(\frac{E_0}{Q_0}\Bigr)
   =
   \Bigl(\frac{dE_0}{d\omega_0}\Bigr)\frac{1}{Q_0}
   -
   \frac{E_0}{Q_0^2}\,\Bigl(\frac{dQ_0}{d\omega_0}\Bigr)
   >
   0 \>.
\end{equation}
This sign can equally well be read off from $Q_0$ alone. The soliton
$\Phi_0(\omega_0)$ extremizes $F[\Phi;\omega_0]\equiv
E[\Phi]-\omega_0Q[\Phi]$ at fixed $\omega_0$ -- this extremization is
exactly the static field equation, Eq.~\eqref{e:EOM0-UVW-sigma}, with
$\omega_0$ the Lagrange multiplier enforcing the charge. Writing
$V(\omega_0)\equiv F[\Phi_0(\omega_0);\omega_0]
=E_0(\omega_0)-\omega_0Q_0(\omega_0)$, the envelope theorem gives
$dV/d\omega_0=\partial F/\partial\omega_0|_{\Phi_0}=-Q_0$: the
implicit dependence of $V$ on $\omega_0$ through the shape of
$\Phi_0(\omega_0)$ drops out, since $\Phi_0$ extremizes $F$ at every
$\omega_0$ individually. Expanding $dV/d\omega_0$ directly instead
gives
\begin{equation}\label{e:E0Q0-envelope}
   \frac{dE_0}{d\omega_0} = \omega_0\,\frac{dQ_0}{d\omega_0} \>,
\end{equation}
so that, by the quotient rule,
\begin{equation}\label{e:EQratio-VK}
   \frac{d}{d\omega_0}\Bigl(\frac{E_0}{Q_0}\Bigr)
   =
   \frac{1}{Q_0}\,\frac{dQ_0}{d\omega_0}\,\Bigl(\omega_0-\frac{E_0}{Q_0}\Bigr) \>.
\end{equation}
Since $Q_0>0$ always, and we find $E_0/Q_0>\omega_0$ numerically
throughout the domain of convergence in every case checked, the sign
of $d(E_0/Q_0)/d\omega_0$ is exactly opposite that of $dQ_0/d\omega_0$:
the familiar statement $dQ_0/d\omega_0<0\Leftrightarrow$ stable is
equivalent to $E_0/Q_0$ increasing monotonically in $\omega_0$. Either
derivative can be evaluated numerically, under the integral sign, from
Eq.~\eqref{e:Q0-y} or Eq.~\eqref{e:E0overQ0}; we apply this criterion
to the SS-PP and VV-PP cases in Secs.~\ref{ss:SSPPstability} and
\ref{ss:VVPPstability} below.

\subsection{\label{ss:BdG}Bogoliubov-de~Gennes eigenvalue problem}

From Sec.~\ref{ss:FirstOrder}, the first-order equation is
\begin{align}
   &\Bigl(\, \xi \, \partial_x + \sigma_3 \bigl[\, m - g_s V^{\kappa} \,\bigr]
   - \bigl[\, \omega_0 + g_v U^{\kappa} \,\bigr] - g_p \, W_{\kappa} \, \sigma_1 \,\Bigr) \Phi_1
   \notag \\
   &\quad
   - \, g_s \kappa \, V^{\kappa-1} \,
   \bigl[\, ( \bar\Phi_0 \Phi_1 ) + ( \bar\Phi_1 \Phi_0 ) \,\bigr] \, \sigma_3 \Phi_0
   \notag \\
   &\quad
   - \, g_v U^{\kappa-1} \Bigl\{\,
   \kappa \, \bigl[\, ( \Phi_0^{\dagger} \Phi_1 ) + ( \Phi_1^{\dagger} \Phi_0 ) \,\bigr] \, \Phi_0
   +
   \bigl[\, ( \Phi_0^{\dagger} \xi \Phi_1 ) + ( \Phi_1^{\dagger} \xi \Phi_0 ) \,\bigr] \, \xi\Phi_0
   \,\Bigr\}
   \notag \\
   &\quad
   - \, g_p \kappa \, \lvert W\rvert^{\kappa-1} \,
   \bigl[\, ( \bar\Phi_0 \xi \Phi_1 ) + ( \bar\Phi_1 \xi \Phi_0 ) \,\bigr] \, \sigma_1 \Phi_0
   =
   \rmi \, \partial_t \Phi_1 \>,
   \label{e:Phi1-xi-eom}
\end{align}
with $V,U,W$ as in Eq.~\eqref{e:UVW-Rtheta}, evaluated on $\Phi_0$.

Each bracket in Eq.~\eqref{e:Phi1-xi-eom} converts into the matrices
$M_0,M_s,M_p,M_v$ already defined in Eq.~\eqref{e:AllMatrices}:
\begin{subequations}\label{e:Phi1-M-terms}
\begin{align}
   \bigl[\, ( \bar\Phi_0 \Phi_1 ) + ( \bar\Phi_1 \Phi_0 ) \,\bigr] \, \sigma_3 \Phi_0
   &=
   M_s \, ( \Phi_1 + \Phi_1^{\ast} ) \>,
   \label{e:Phi1-M-terms-a} \\
   \bigl[\, ( \Phi_0^{\dagger} \Phi_1 ) + ( \Phi_1^{\dagger} \Phi_0 ) \,\bigr] \, \Phi_0
   &=
   M_0 \, ( \Phi_1 + \Phi_1^{\ast} ) \>,
   \label{e:Phi1-M-terms-b} \\
   \bigl[\, ( \Phi_0^{\dagger} \xi \Phi_1 ) + ( \Phi_1^{\dagger} \xi \Phi_0 ) \,\bigr] \, \xi\Phi_0
   &=
   - M_v \, ( \Phi_1 - \Phi_1^{\ast} ) \>,
   \label{e:Phi1-M-terms-c} \\
   \bigl[\, ( \bar\Phi_0 \xi \Phi_1 ) + ( \bar\Phi_1 \xi \Phi_0 ) \,\bigr] \, \sigma_1 \Phi_0
   &=
   M_p \, ( \Phi_1 + \Phi_1^{\ast} ) \>.
   \label{e:Phi1-M-terms-d}
\end{align}
\end{subequations}
The symmetric dressing matrices $\sigma_3,\sigma_1$ (and the trivial
dressing of the second identity) pair with $\Phi_1+\Phi_1^{\ast}$,
while the antisymmetric $\xi$-dressing of the third identity pairs
instead with $\Phi_1-\Phi_1^{\ast}$.

Putting these results into Eq.~\eqref{e:Phi1-xi-eom} casts it as
\begin{equation}\label{e:HM-form-eq}
   H(x)\,\Phi_1(x,t) + M(x)\,\Phi_1^{\ast}(x,t)
   =
   \rmi \, \partial_t \Phi_1(x,t) \>,
\end{equation}
where $H(x)$ and $M(x)$ are both real and Hermitian,
\begin{subequations}\label{e:HM-defs}
\begin{align}
   H(x)
   &=
   \xi \, \partial_x + \sigma_3 \bigl[\, m - g_s V^{\kappa} \,\bigr]
   - \bigl[\, \omega_0 + g_v U^{\kappa} \,\bigr] - g_p \, W_{\kappa} \, \sigma_1
   \notag \\
   &\quad
   - \, g_s \kappa \, V^{\kappa-1} \, M_s
   - g_v \kappa \, U^{\kappa-1} \, M_0
   + g_v U^{\kappa-1} \, M_v
   - g_p \kappa \, \lvert W\rvert^{\kappa-1} \, M_p \>,
   \label{e:HH-def} \\
   M(x)
   &=
   - \, g_s \kappa \, V^{\kappa-1} \, M_s
   - g_v \kappa \, U^{\kappa-1} \, M_0
   - g_v U^{\kappa-1} \, M_v
   - g_p \kappa \, \lvert W\rvert^{\kappa-1} \, M_p \>.
   \label{e:MM-def}
\end{align}
\end{subequations}
Since $H,M$ are real, adjoining the complex conjugate of
Eq.~\eqref{e:HM-form-eq} as a second row casts it as the coupled
matrix equation
\begin{equation}\label{e:Phi1-Phi1ast-matrix}
   \begin{pmatrix} H & M \\ M & H \end{pmatrix}
   \begin{pmatrix} \Phi_1(x,t) \\ \Phi_1^{\ast}(x,t) \end{pmatrix}
   =
   \begin{pmatrix} \rmi \partial_t & 0 \\ 0 & -\rmi \partial_t \end{pmatrix}
   \begin{pmatrix} \Phi_1(x,t) \\ \Phi_1^{\ast}(x,t) \end{pmatrix} \>.
\end{equation}
Seeking normal-mode solutions
\begin{subequations}\label{e:Phi1-normal-mode}
\begin{align}
   \Phi_1(x,t)
   &=
   F_n(x) \, \rme^{-\rmi \nu_n t}
   +
   G_n^{\ast}(x) \, \rme^{+\rmi \nu_n^{\ast} t} \>,
   \label{e:Phi1-normal-mode-a} \\
   \Phi_1^{\ast}(x,t)
   &=
   F_n^{\ast}(x) \, \rme^{+\rmi \nu_n^{\ast} t}
   +
   G_n(x) \, \rme^{-\rmi \nu_n t} \>,
   \label{e:Phi1-normal-mode-b}
\end{align}
\end{subequations}
and matching the coefficients of $\rme^{-\rmi\nu_n t}$ and
$\rme^{+\rmi\nu_n^{\ast} t}$ separately in each row of
Eq.~\eqref{e:Phi1-Phi1ast-matrix} gives the Bogoliubov-de~Gennes (B-G)
eigenvalue problem \cite{Bogolyubov-1947,deGennes-1966}
\begin{equation}\label{e:eigen-problem}
   \begin{pmatrix} H(x) & M(x) \\ -M(x) & -H(x) \end{pmatrix}
   \begin{pmatrix} F_n(x) \\ G_n(x) \end{pmatrix}
   =
   \nu_n \,
   \begin{pmatrix} F_n(x) \\ G_n(x) \end{pmatrix} \>.
\end{equation}
Note that $\nu_n$ is a frequency measured in the frame already
co-rotating with the background, Eq.~\eqref{e:PsiExpand}'s
overall factor $\rme^{-\rmi\omega_0 t}$ -- not the lab-frame frequency
of the corresponding piece of $\Psi$ itself. Reading $\nu_n$'s
$F_n$-component off the full field via Eqs.~\eqref{e:PsiExpand}
and \eqref{e:Phi1-normal-mode-a}, $\Psi\supset \varepsilon
F_n(x)\,\rme^{-\rmi(\omega_0+\nu_n)t}$: the lab-frame frequency is
$\omega=\omega_0+\nu_n$ (and $\omega_0-\nu_n$ for the conjugate
$G_n^{\ast}$ piece), not $\nu_n$ alone. This is also why linearizing
the full, lab-frame nonlinear equation directly -- as in
Secs.~\ref{ss:TimeEvolution} and \ref{ss:TimeEvolutionVVPP}'s Jacobian
checks -- reproduces $\lambda_n=-\rmi\nu_n$ in the real part but
with an extra $-\rmi\omega_0$ appearing alongside it in the imaginary
part.
Properties of its eigenvalues and eigenfunctions are well known
\cite{BoussaidComech-book-2019,Berkolaiko_2015}:
\begin{enumerate}[noitemsep]
\item There are two zero-eigenvalue ($\nu_n=0$) solutions, one for
each continuous symmetry (phase and translation) of the static field
equations. At $\nu_n=0$, Eq.~\eqref{e:Phi1-normal-mode} reduces to the
time-independent $\Phi_1(x)=F_n(x)+G_n^{\ast}(x)$.
\begin{enumerate}[noitemsep]
\item[(i)] $F_n=G_n^{\ast}=\tfrac{\rmi}{2}\Phi_0(x)$ requires
$[H(x)-M(x)]\Phi_0=0$. From Eqs.~\eqref{e:HH-def}--\eqref{e:MM-def},
\begin{equation*}
   H(x) - M(x)
   =
   \xi \, \partial_x + \sigma_3 ( m - g_s V^{\kappa} )
   - ( \omega_0 + g_v U^{\kappa} ) - g_p \, W_{\kappa} \, \sigma_1
   + 2 g_v U^{\kappa-1} \, M_v \>,
\end{equation*}
so $[H-M]\Phi_0$ is the sum of the left side of
Eq.~\eqref{e:EOM0-UVW-sigma}, which vanishes since $\Phi_0$ solves it,
and $2g_vU^{\kappa-1}M_v\Phi_0$, which also vanishes since
$M_v\Phi_0=-\xi\Phi_0(\Phi_0^{T}\xi\Phi_0)=0$
($\Phi_0^{T}\xi\Phi_0=0$ identically for the antisymmetric $\xi$).
This phase-type zero mode,
\begin{equation}\label{e:phase-zero-mode}
   \Phi_{1,0}(x) \equiv \frac{\rmi}{2}
   \begin{pmatrix} \Phi_0(x) \\ -\Phi_0(x) \end{pmatrix} \>,
\end{equation}
gives $\Phi_1(x,t)=\rmi\,\Phi_0(x)$ via Eq.~\eqref{e:Phi1-normal-mode}
-- the expected infinitesimal $\mathrm{U}(1)$ phase rotation of the
soliton.
\item[(ii)] $F_n=G_n=\tfrac12\Phi_0'(x)$ [real] requires
$[H(x)+M(x)]\Phi_0'=0$. Here
\begin{align*}
   H(x) + M(x)
   &=
   \xi \, \partial_x + \sigma_3 ( m - g_s V^{\kappa} )
   - ( \omega_0 + g_v U^{\kappa} ) - g_p \, W_{\kappa} \, \sigma_1
   \\
   &\quad
   - 2 g_s \kappa V^{\kappa-1} M_s
   - 2 g_v \kappa U^{\kappa-1} M_0
   - 2 g_p \kappa \lvert W \rvert^{\kappa-1} M_p \>,
\end{align*}
the $M_v$ terms of $H$ and $M$ having cancelled. This combination is
exactly the $x$-derivative of Eq.~\eqref{e:EOM0-UVW-sigma}
[using $(\Phi_0^{T}\sigma_3\Phi_0')\sigma_3\Phi_0=M_s\Phi_0'$,
$(\Phi_0^{T}\Phi_0')\Phi_0=M_0\Phi_0'$, and
$(\Phi_0^{T}\sigma_1\Phi_0')\sigma_1\Phi_0=M_p\Phi_0'$ to rewrite the
chain-rule terms], so $[H+M]\Phi_0'=0$ since $\Phi_0$ solves
Eq.~\eqref{e:EOM0-UVW-sigma} for every $x$. This translation-type zero
mode,
\begin{equation}\label{e:trans-zero-mode}
   \Phi_{2,0}(x) \equiv \frac12
   \begin{pmatrix} \Phi_0'(x) \\ \Phi_0'(x) \end{pmatrix} \>,
\end{equation}
gives $\Phi_1(x,t)=\Phi_0'(x)$ -- the expected infinitesimal spatial
translation of the soliton.
\end{enumerate}
$\Phi_{1,0}$ and $\Phi_{2,0}$ span the two-dimensional degenerate
zero-eigenspace of Eq.~\eqref{e:eigen-problem}.
\item For every eigenvalue $\nu_n$ there is a second eigenvalue
$-\nu_n^{\ast}$: taking the complex conjugate of
Eq.~\eqref{e:eigen-problem} and using that $H,M$ are real shows
$(G_n^{\ast},F_n^{\ast})$ is itself an eigenvector, with eigenvalue
$-\nu_n^{\ast}$. (a) If $n=m$ and $\int(\lvert F_n\rvert^2-\lvert
G_n\rvert^2)\,dx\neq0$, this pairing argument shows $\nu_n$ is real; a
vanishing norm signals a genuinely complex eigenvalue pair, i.e.\ an
instability. (b) Labeling the paired mode $-n$ (so
$\nu_{-n}=-\nu_n$, $(F_{-n},G_{-n})=(G_n^{\ast},F_n^{\ast})$), the
real-eigenvalue eigenfunctions may be normalized as
\begin{equation}\label{e:orthonormality}
   \int \! dx \, \bigl(\, F_m^{\dagger} F_n - G_m^{\dagger} G_n \,\bigr)
   = \frac{n}{\lvert n\rvert} \, \delta_{n,m} \>.
\end{equation}
\end{enumerate}
Both zero modes are orthogonal, in the sense of
Eq.~\eqref{e:orthonormality}, to every $n>0$ eigenmode, since
$\nu_{1,0}=\nu_{2,0}=0\neq\nu_n$. As in the Bogoliubov approximation,
the physical fluctuation $\Phi_1$ is required to carry no component
along the zero modes -- an infinitesimal phase rotation or
translation of the background $\Phi_0$ is not an independent
perturbation -- so the general solution is the superposition of
$n>0$ pairs alone,
\begin{equation}\label{e:general-solution}
   \begin{pmatrix} \Phi_1(x,t) \\ \Phi_1^{\ast}(x,t) \end{pmatrix}
   =
   \sum_{n>0}
   \Biggl\{
   C_n
   \begin{pmatrix} F_n(x) \\ G_n(x) \end{pmatrix}
   \rme^{-\rmi \nu_n t}
   +
   C_n^{\ast}
   \begin{pmatrix} G_n^{\ast}(x) \\ F_n^{\ast}(x) \end{pmatrix}
   \rme^{+\rmi \nu_n^{\ast} t}
   \Biggr\} \>,
\end{equation}
with coefficients fixed, via Eq.~\eqref{e:orthonormality}, by the
initial fluctuation $\Phi_1(x,0)$ alone (not by $\Phi_0$, which is a
separate, zeroth-order piece of Eq.~\eqref{e:PsiExpand}):
\begin{equation}\label{e:Cn-projection}
   C_n = \int \! dx \, \bigl[\, F_n^{\dagger}(x)\, \Phi_1(x,0)
   - G_n^{\dagger}(x)\, \Phi_1^{\ast}(x,0) \,\bigr] \>, \qquad n>0 \>.
\end{equation}
The soliton is linearly stable if and only if every $\nu_n$ in
Eq.~\eqref{e:eigen-problem}, beyond the two zero modes, is real.

\subsection{\label{ss:AltEigen}Alternate eigenvalue equation}

The complex eigenvalue equation Eq.~\eqref{e:eigen-problem} can be
brought to an equivalent real form by the change of variables
\begin{equation}\label{e:XY-change-of-variables}
   F_n(x) = X_n(x) + \rmi \, Y_n(x) \>, \qquad
   G_n(x) = X_n(x) - \rmi \, Y_n(x) \>.
\end{equation}
Substituting into the two rows of Eq.~\eqref{e:eigen-problem} and
adding/subtracting decouples them: adding gives
$2\rmi(H-M)Y_n=2\nu_nX_n$, while subtracting gives
$2(H+M)X_n=2\rmi\nu_nY_n$, i.e.,
\begin{equation}\label{e:XY-matrix-eigen}
   \begin{pmatrix} 0 & L_{-}(x) \\ -L_{+}(x) & 0 \end{pmatrix}
   \begin{pmatrix} X_n(x) \\ Y_n(x) \end{pmatrix}
   =
   \lambda_n \,
   \begin{pmatrix} X_n(x) \\ Y_n(x) \end{pmatrix} \>, \qquad
   \lambda_n \equiv - \rmi \, \nu_n \>,
\end{equation}
where
\begin{subequations}\label{e:LXLY-def}
\begin{align}
   L_{+}(x)
   &\equiv
   H(x) + M(x)
   \notag \\
   &=
   \xi \, \partial_x + \sigma_3 \bigl[\, m - g_s V^{\kappa} \,\bigr]
   - \bigl[\, \omega_0 + g_v U^{\kappa} \,\bigr] - g_p \, W_{\kappa} \, \sigma_1
   \notag \\
   &\quad
   - 2 g_s \kappa \, V^{\kappa-1} \, M_s
   - 2 g_v \kappa \, U^{\kappa-1} \, M_0
   - 2 g_p \kappa \, \lvert W\rvert^{\kappa-1} \, M_p \>,
   \label{e:LX-explicit} \\
   L_{-}(x)
   &\equiv
   H(x) - M(x)
   \notag \\
   &=
   \xi \, \partial_x + \sigma_3 \bigl[\, m - g_s V^{\kappa} \,\bigr]
   - \bigl[\, \omega_0 + g_v U^{\kappa} \,\bigr] - g_p \, W_{\kappa} \, \sigma_1
   + 2 g_v \, U^{\kappa-1} \, M_v \>,
   \label{e:LY-explicit}
\end{align}
\end{subequations}
the second equality in each line following from Eq.~\eqref{e:HM-defs},
since the $M_v$ contribution cancels in $L_{+}$ while the
$M_s,M_0,M_p$ contributions cancel in $L_{-}$. $L_+$ and $L_-$ are
each individually Hermitian, with their own real-eigenvalue problem
$L_{\pm}(x)\phi_n^{\pm}(x)=\mu_n^{\pm}\phi_n^{\pm}(x)$, independent of
the coupled problem Eq.~\eqref{e:XY-matrix-eigen}. Eliminating one of
$X_n,Y_n$ instead decouples Eq.~\eqref{e:XY-matrix-eigen} into a pair
of second-order equations,
$L_{-}L_{+}X_n=-\lambda_n^2X_n$, $L_{+}L_{-}Y_n=-\lambda_n^2Y_n$;
because $L_+,L_-$ need not commute, $L_-L_+$ and $L_+L_-$ need not be
Hermitian, so $\lambda_n$ need not be purely real or purely imaginary.

The two zero modes of Sec.~\ref{ss:BdG} translate directly into this
real form: $L_+$ of Eq.~\eqref{e:LX-explicit} is exactly $H+M$ of
item~1(ii) there, so $\Phi_0'(x)$ is its zero mode (translation); $L_-$
of Eq.~\eqref{e:LY-explicit} is exactly $H-M$ of item~1(i), so
$\Phi_0(x)$ is its zero mode (phase):
\begin{equation}\label{e:XY-zero-modes}
   \begin{pmatrix} X_{1,0}(x) \\ Y_{1,0}(x) \end{pmatrix}
   =
   \begin{pmatrix} 0 \\ \Phi_0(x) \end{pmatrix} \>, \qquad
   \begin{pmatrix} X_{2,0}(x) \\ Y_{2,0}(x) \end{pmatrix}
   =
   \begin{pmatrix} \Phi_0'(x) \\ 0 \end{pmatrix} \>.
\end{equation}
This real $(X_n,Y_n)$ form -- rather than the complex $(F_n,G_n)$ form
of Sec.~\ref{ss:BdG} -- is the one we discretize numerically in
Secs.~\ref{ss:SSPPstability} and \ref{ss:VVPPstability} below, via the
Chebyshev spectral method described next.

\subsection{\label{ss:ChebyshevMethod}Chebyshev spectral discretization}

Equation~\eqref{e:XY-matrix-eigen} is a first-order linear operator
eigenvalue problem on the whole real line, with $\Phi_0(x)\to0$ as
$x\to\pm\infty$ the only boundary condition. We solve it by mapping
$x\in(-\infty,\infty)$ onto a finite Chebyshev grid via the algebraic
map
\begin{equation}\label{e:algebraic-map}
   x = \frac{Lt}{\sqrt{1-t^2}} \>, \qquad t\in(-1,1) \>,
\end{equation}
with $L$ a length scale chosen to resolve the soliton's actual width
(set by $\beta$, not $\kappa\beta$; a soliton with a narrower core
needs a smaller $L$ for a given $N$). On the $N+1$ Chebyshev-Gauss-Lobatto
nodes $t_j=\cos(j\pi/N)$, $j=0,\dots,N$, the standard Chebyshev
differentiation matrix is
\begin{equation}\label{e:cheb-matrix}
   D^{t}_{jk} = \frac{c_j}{c_k}\,\frac{1}{t_j-t_k} \ (j\neq k) \>,
   \qquad
   D^{t}_{jj} = -\sum_{k\neq j} D^{t}_{jk} \>,
\end{equation}
with $c_0=c_N=2$, $c_j=1$ otherwise, and $(-1)^j$ absorbed into each
$c_j$; this is the well-known dense, spectrally accurate
differentiation matrix for polynomial interpolants on the
Chebyshev-Gauss-Lobatto grid. The chain rule then gives
$D^x=\mathrm{diag}(dt/dx)\,D^t$, with $dt/dx$ read off from
Eq.~\eqref{e:algebraic-map}. The two endpoints $t=0,N$ (i.e.\
$x=\pm\infty$) are dropped from the grid entirely -- not merely set to
zero -- which enforces $\Phi_0(x)\to0$ there automatically, exactly
the boundary condition the problem requires. Evaluating
$L_{\pm}(x)=\xi\,\partial_x+(\text{local }2\times2\text{ matrix})$ of
Eq.~\eqref{e:LXLY-def} at each interior node, with $D^x$ in place of
$\partial_x$, gives a single dense $2N\times2N$ (real) matrix per
operator $L_+,L_-$; stacking them as in Eq.~\eqref{e:XY-matrix-eigen}
and diagonalizing directly (a single dense eigenvalue solve) returns
the full discrete spectrum at once, with no iterative search needed.

Most of that spectrum is a discretization of the problem's essential
(continuum) spectrum, $(-\infty,-m-\omega_0]\cup[m-\omega_0,\infty)$,
and is unphysical. We retain only eigenvalues that are (i) spatially
localized, with more than $80\%$ of their $L^2$ norm within a fixed
core region (typically $|x|<3$--$4$), and (ii) stable under
independent changes of resolution, matching to better than $0.5\%$
between two separate choices of $(N,L)$ (e.g.\ $N$ and $N+100$) --
genuine eigenvalues converge under refinement, while spurious
continuum artifacts drift or appear only at a single resolution. Both
criteria were used throughout Secs.~\ref{ss:SSPPstability} and
\ref{ss:VVPPstability} below whenever we report a growth rate as
``genuine,'' and it is exactly this second criterion that let us
correctly reject the non-converging spurious real pair found in the
VV-PP stability checks there.

\begin{figure}[t]
   \centering
   \includegraphics[width=\linewidth]{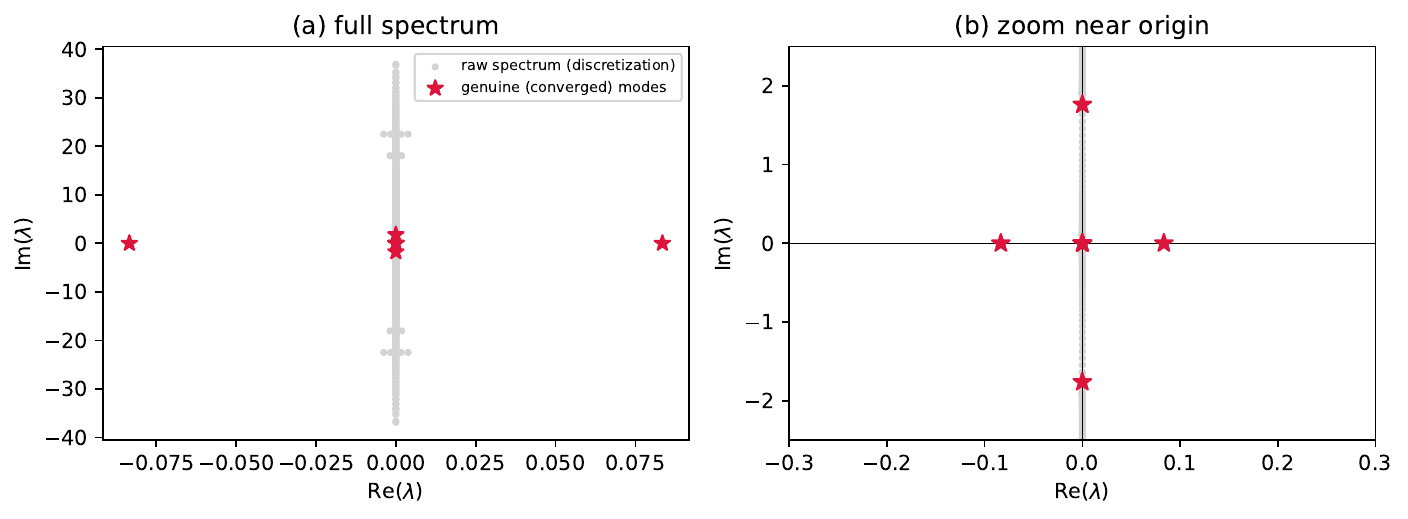}
   \caption{\label{f:SSPPSpectrum}The Bogoliubov-de~Gennes spectrum for
   the SS-PP example of Fig.~\ref{f:SSPPUnstableEigenmode}
   ($\kappa=3$, $q=1$, $r=-0.95$, $\omega_0=0.90$), at $N=300$,
   $L=8$: (a) the full spectrum, raw eigenvalues (gray) against the
   eight that converge under refinement and are spatially localized
   (red stars); (b) a zoom near the origin showing the two zero modes,
   the unstable pair, and the oscillatory pair.}
\end{figure}

A caveat on scope: wherever Secs.~\ref{ss:SSPPstability} and
\ref{ss:VVPPstability} report scanning for a ``genuine growing mode,''
we filtered specifically for $\mathrm{Re}(\lambda)>10^{-3}$, following
a companion paper's \cite{Cooper-Jacobs-Khare-2026} own division of
the question into a
search for real/growing instabilities (its Secs.~V.A, VI, VII.A) and a
separate, explicit search for oscillatory (complex) content (its
Secs.~V.B, VII.B). That filtered search is not an exhaustive catalog
of the discrete spectrum: genuinely stable, purely oscillatory bound
states ($\mathrm{Re}(\lambda)=0$, as opposed to the growing or
decaying $\mathrm{Re}(\lambda)\neq0$ pairs we filter for) can and do
exist in the gap in addition to the two zero modes, and we note them
explicitly wherever a broader, unfiltered scan turned one up.
Figure~\ref{f:SSPPSpectrum} shows what this looks like in practice for
the SS-PP example of Sec.~\ref{ss:SSPPstability}
($\kappa=3$, $q=1$, $r=-0.95$, $\omega_0=0.90$): the full raw spectrum
at a single resolution, with the eight eigenvalues that survive both
filters -- the doubly-degenerate zero mode, the unstable pair
$\lambda=\pm0.0834$, and the oscillatory pair $\nu=\pm1.763\,\rmi$ --
picked out against the discretized continuum. Every eigenvalue here,
genuine or not, falls exactly on one axis or the other: unlike the
genuinely complex quartet found on the negative-frequency VV branch of
the companion paper \cite{Cooper-Jacobs-Khare-2026}, no off-axis
content appears anywhere in this example.

\subsubsection{\label{ss:SLACcompare}The SLAC derivative and fermion doubling}

A natural question is whether the Chebyshev method's eigenvalues are
physical or an artifact of its own, non-uniform discretization. We
address this the way lattice field theorists address the analogous
question for a uniform grid: by comparing against a discretization
scheme designed specifically to avoid \emph{fermion doubling}, the
spurious low-energy mode that an ordinary local finite-difference
stencil manufactures from the first-derivative operator's dispersion
relation folding back at the lattice's Brillouin-zone edge. An
ordinary centered difference, $(\partial_xf)_j\approx(f_{j+1}-f_{j-1})/2a$
on a uniform lattice of spacing $a$, has discrete dispersion
$\omega(k)=\sin(ka)/a$: correct at small $ka$, but folding back to
zero at $k=\pm\pi/a$ rather than growing linearly, which manufactures
a doubler mode degenerate with the physical one. The SLAC derivative
of Drell, Weinstein, and Yankielowicz
\cite{Drell-Weinstein-Yankielowicz-1976} removes this fold-back by
replacing the sparse local stencil with the dense matrix
\begin{equation}\label{e:slac-derivative}
   D^{\rm SLAC}_{jk} =
   \begin{cases}
      0 \>, & j=k \>, \\[4pt]
      \dfrac{\pi}{N_{\rm latt}a}\,\dfrac{(-1)^{j-k}}{
      \sin\!\big(\pi(j-k)/N_{\rm latt}\big)} \>, & j\neq k \>,
   \end{cases}
\end{equation}
for $N_{\rm latt}$ points on a periodic ring of spacing $a$;
Eq.~\eqref{e:slac-derivative} reproduces $\omega(k)=k$ exactly over
the whole Brillouin zone, exactly the same non-locality the Chebyshev
method trades for dispersion accuracy, just on a uniform periodic
grid rather than a non-uniform open one. We follow the same
Chebyshev-versus-SLAC comparison used for the related massive
Gross-Neveu model, where it was part of a more complete survey of the
Vakhitov-Kolokolov criterion over a wide range of couplings and
nonlinearities \cite{Dawson-GN-2026}.

We already carried out exactly this cross-check, directly on our own
soliton and our own growth rates rather than a separate toy problem:
Secs.~\ref{ss:TimeEvolution} and \ref{ss:TimeEvolutionVVPP} evolve the
full nonlinear equation with the SLAC derivative of
Eq.~\eqref{e:slac-derivative}, starting from the Chebyshev-derived
unstable eigenmode in each case, and the resulting growth rate agrees
with the Chebyshev value to four significant figures throughout both
runs (the ratio of the two methods' $\|\rho(t)-\rho(0)\|_2$ curves is
constant at $2.8999$--$2.9003$ for SS-PP). Two discretizations with
nothing in common beyond both avoiding fermion doubling -- one
non-uniform and open, the other uniform and periodic -- agreeing on
every genuine growth rate checked is the same confirmation lattice
field theorists use Eq.~\eqref{e:slac-derivative} for: it rules out a
discretization-specific artifact as the origin of the eigenvalues we
report as physical throughout Sec.~\ref{s:Stability}.

\subsection{\label{ss:SSPPstability}Specialization to SS-PP}

For the SS-PP case ($g_v=0$), direct numerical evaluation of
$E_0/Q_0(\omega_0)$ at $q=1$, $r=-0.95$ shows monotonic increase
throughout $\omega_{\rm exist}<\omega_0<m$ for $\kappa\lesssim2$, but
an interior maximum -- hence a VK-unstable window near $\omega_0\to
m$ -- appearing for $\kappa\gtrsim2.2$ (e.g.\ at $\kappa=3$ the
maximum occurs at $\omega_0\approx0.88\,m$). This matches, both
qualitatively and at the same critical $\kappa\approx2$, the general
SS-VV-PP finding of Shao, Quintero, Mertens, Cooper, Khare, and Saxena
\cite{PhysRevE.90.032915} that $E_0/Q_0$ develops
a negative slope near $\omega_0\to m$ once $\kappa>2$, signaling
instability there.

\begin{figure}[t]
   \centering
   \includegraphics[width=\linewidth]{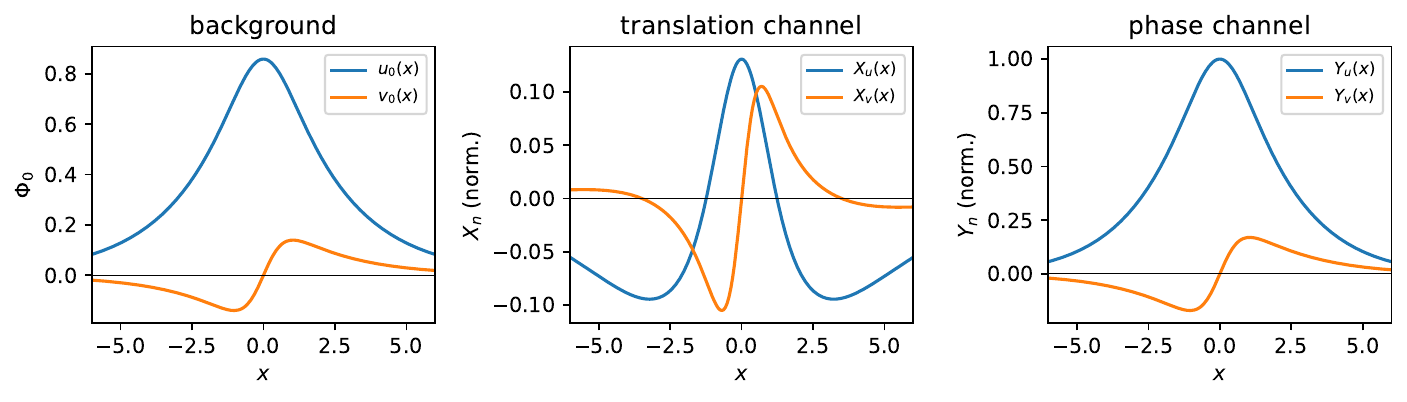}
   \caption{\label{f:SSPPUnstableEigenmode}SS-PP, $\kappa=3$, $q=1$,
   $r=-0.95$, $\omega_0=0.90$: background $\Phi_0=(u_0,v_0)$ (left) and
   the genuine unstable Bogoliubov-de~Gennes eigenmode
   $(X_n,Y_n)$ (center, right; each normalized to unit peak
   height), with growth rate $\lambda_n=0.083$.}
\end{figure}

We independently confirmed this onset by directly diagonalizing the
full Bogoliubov-de~Gennes problem of Sec.~\ref{ss:BdG}, using the
Chebyshev spectral discretization of Sec.~\ref{ss:ChebyshevMethod}
applied to Sec.~\ref{ss:AltEigen}'s real $(X_n,Y_n)$ form. Scanning
$\omega_0$ at $\kappa=3$, $q=1$, $r=-0.95$,
we find no genuine (spatially localized, resolution-stable) growing
mode at $\omega_0=0.80,0.85$, and a genuine real growth rate
$\lambda=0.083,0.105,0.095$ at $\omega_0=0.90,0.93,0.95$ respectively
-- the instability turning on exactly where the $E_0/Q_0$ criterion
above predicts. Figure~\ref{f:SSPPUnstableEigenmode} shows the
background $\Phi_0$ together with the unstable eigenmode
$(X_n,Y_n)$ at $\omega_0=0.90$ ($\lambda=0.083$); via
Eq.~\eqref{e:XY-change-of-variables}, $F_n=X_n+\rmi Y_n$ is real-$\lambda_n$
here, so the physical perturbation is simply
$\Phi_1(x,0)\propto X_n(x)+\rmi Y_n(x)$. Relaxing the growing-mode
filter at $\omega_0=0.90$, the genuine (resolution-stable, localized)
part of the spectrum consists of exactly four modes, not two: the two
zero modes, the unstable pair $\lambda=\pm0.0834$ already discussed,
and one additional discrete, purely oscillatory bound state at
$\nu=\pm1.763\,\rmi$ -- well inside the spectral gap
$(-m-\omega_0,m-\omega_0)=(-1.9,0.1)$ and confirmed box-independent
across $L=8$--$12$. This mode is stable
($\mathrm{Re}(\lambda)=0$ to numerical precision) and plays no role in
the instability; we return to its VV-PP counterpart in
Sec.~\ref{ss:VVPPstability}.

\subsubsection{\label{ss:TimeEvolution}Nonlinear time evolution}

As a final, direct check of the instability, we evolve the full
nonlinear Eq.~\eqref{e:EOM} (not its linearization) forward from the
perturbed initial condition
\begin{equation}
   \Psi(x,0) = \Phi_0(x) + \varepsilon\,\Phi_1(x,0) \>,
   \qquad
   \Phi_1(x,0) = X_n(x) + \rmi\,Y_n(x) \>,
\end{equation}
with $(X_n,Y_n)$ the genuine unstable eigenmode of
Fig.~\ref{f:SSPPUnstableEigenmode} ($\kappa=3$, $q=1$, $r=-0.95$,
$\omega_0=0.90$, $\lambda_n=0.0834$) and $\varepsilon$ fixed, following
Cooper, Jacobs, and Khare's companion paper
\cite{Cooper-Jacobs-Khare-2026}, by
$\max_x|\varepsilon\Phi_1(x,0)|=10^{-3}\max_x|\Phi_0(x)|$. This is the
purest possible test case for Eq.~\eqref{e:Cn-projection}: rather than
projecting some generic initial perturbation onto the normal-mode
basis, we construct $\Phi_1(x,0)$ to equal a single eigenmode outright,
so that $C_n=1$ for this mode and $C_m=0$ for every other $m$
trivially, without needing to evaluate the projection integral itself.
We integrate
with fourth-order Runge-Kutta on a periodic lattice using the SLAC
derivative -- exact multiplication by $\rmi k$ in Fourier space, free
of both fermion doubling and the mass shift that local finite-difference
stencils introduce -- rather than the Chebyshev discretization used
above, since the wide, uniform lattice this method needs is also what
exposed a numerical subtlety worth recording: evaluating $R^2(x)$ from
the literal $m\cos(2\theta)-\omega_0$ of Eq.~\eqref{e:Rsq-formula}
loses all precision far in the tails, to the point of giving the wrong
background entirely by $|x|\sim15$ for this $\kappa$; we instead use
the cancellation-free identity already given in
Eq.~\eqref{e:cos-sin-2theta-y} (the same one used for this reason in
a companion paper \cite{Cooper-Jacobs-Khare-2026}), verified against
$\Phi_0(x)$ computed in
50-digit arbitrary precision.

\begin{figure}[t]
   \centering
   \includegraphics[width=\linewidth]{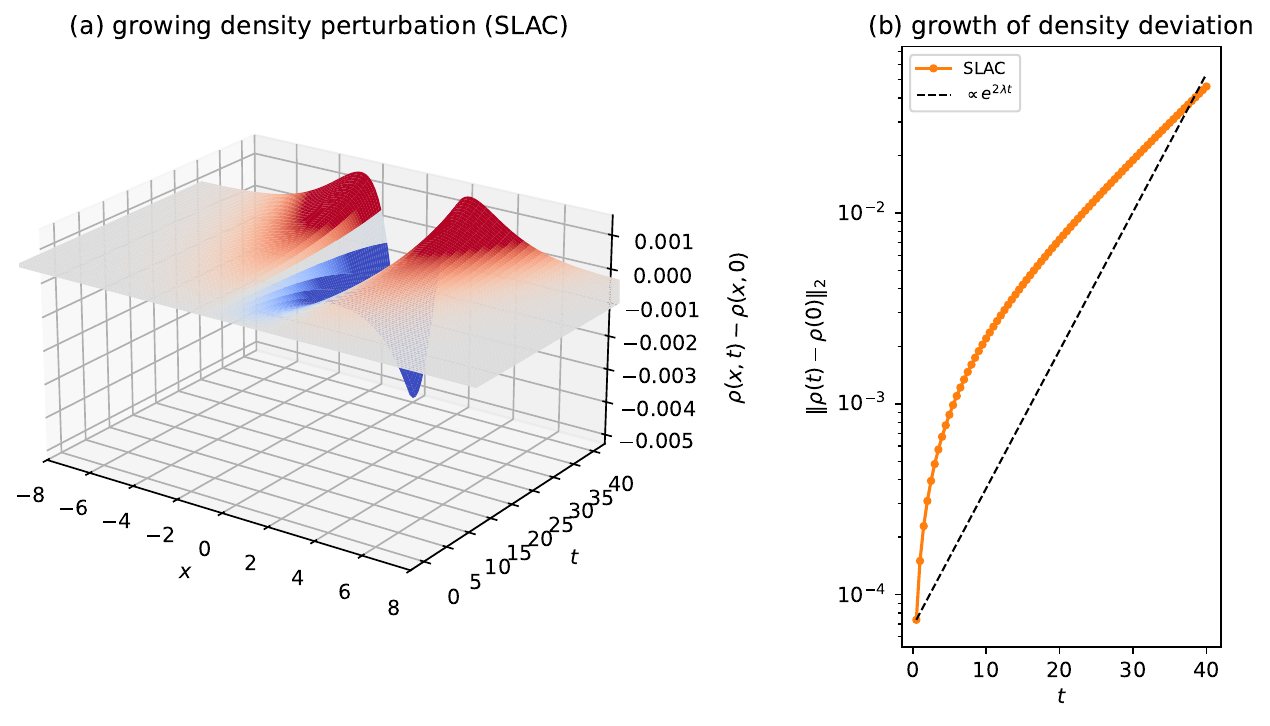}
   \caption{\label{f:SSPPTimeEvolution}Nonlinear time evolution of the
   perturbed SS-PP soliton of Fig.~\ref{f:SSPPUnstableEigenmode}: (a)
   the growing density perturbation $\rho(x,t)-\rho(x,0)$; (b) its
   $L^2$ norm, approaching the predicted asymptotic slope $2\lambda_n$
   (dashed).}
\end{figure}

Figure~\ref{f:SSPPTimeEvolution}(a) shows the growing density
perturbation $\rho(x,t)-\rho(x,0)$ directly; panel (b) shows
$\|\rho(t)-\rho(0)\|_2$ on a log scale, converging to the predicted
asymptotic slope $2\lambda_n$ (the density being quadratic in $\Psi$)
once the transient from the subdominant modes excited by the initial
perturbation has decayed. Charge is conserved to $1.5\times10^{-12}$
over the full run. As a cross-check, we repeated the same evolution on
the original Chebyshev grid -- correcting its much shorter domain's
fixed discretization residual at $\Phi_0$, which does not vanish
exactly the way the SLAC derivative's does, by subtracting it at every
step, a correction of magnitude only $10^{-9}$ once the same
cancellation-free $\Phi_0$ is used -- and found the identical growth
rate: the ratio of the two methods' $\|\rho(t)-\rho(0)\|_2$ curves is
constant to four significant figures ($2.8999$-$2.9003$) over the
entire run, the two methods differing only in how the single shared
initial eigenmode happens to be normalized on their respective grids.

\begin{figure}[t]
   \centering
   \includegraphics[width=\linewidth]{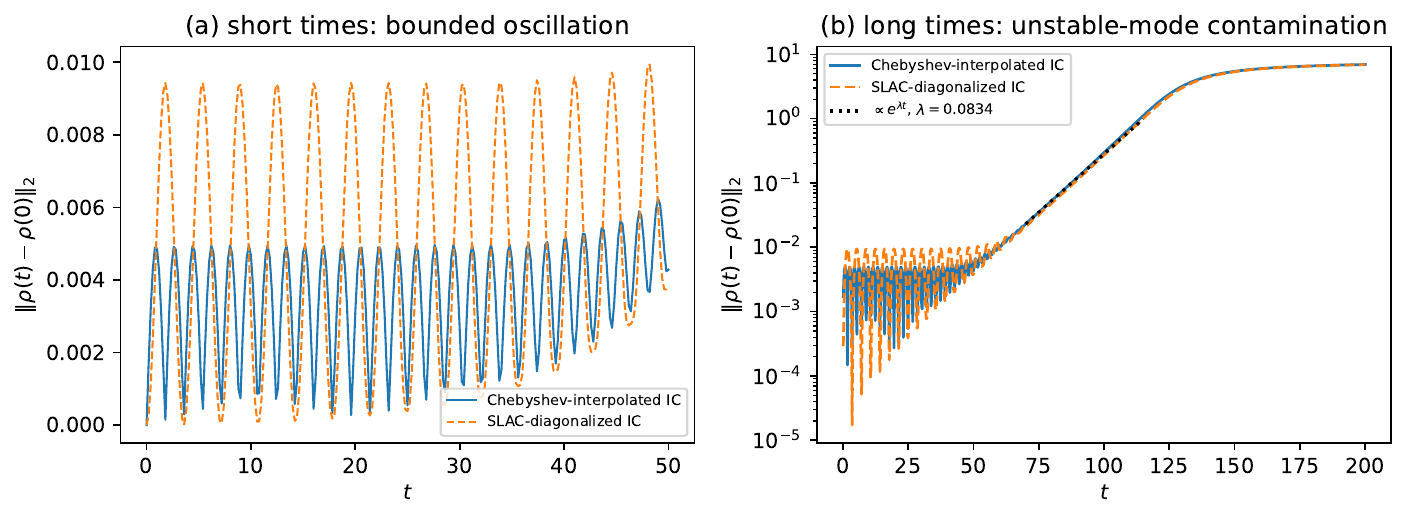}
   \caption{\label{f:SSPPOscillatoryContamination}Nonlinear evolution
   seeded with the stable oscillatory mode ($\nu=1.763$) instead of
   the unstable one: (a) bounded oscillation at short times, from two
   independently constructed initial conditions (Chebyshev-interpolated
   vs.\ directly SLAC-diagonalized); (b) both eventually grow at the
   \emph{unstable} mode's rate $\lambda_n=0.0834$ (dotted) as
   unavoidable roundoff contamination takes over, agreeing closely
   regardless of which grid the eigenvector was computed on.}
\end{figure}

As a complementary check, we repeated this evolution seeding
$\Phi_1(x,0)$ with the \emph{stable} oscillatory mode found above
($\nu=1.763$) instead of the unstable
one, using $\Phi_1(x,0)=F_n(x)+G_n^{\ast}(x)$ [Eq.~\eqref{e:Phi1-normal-mode-a}
at $t=0$] with the same $10^{-3}\max_x|\Phi_0(x)|$ normalization.
Figure~\ref{f:SSPPOscillatoryContamination}(a) shows exactly the
bounded, non-growing oscillation the linear theory predicts over the
first $\sim14$ periods ($t\lesssim50$). Run far longer, however, a
second regime appears: $\|\rho(t)-\rho(0)\|_2$ begins growing
[Fig.~\ref{f:SSPPOscillatoryContamination}(b)] at a rate
($0.086$, fit over $t=70$--$115$) matching the \emph{unstable} mode's
own $\lambda_n=0.0834$, before saturating at the same nonlinear
plateau as Fig.~\ref{f:SSPPTimeEvolution}. This is not a nonlinear
instability of the oscillatory mode itself: constructing
$\Phi_1(x,0)$ numerically cannot avoid leaving an infinitesimal
admixture of the system's genuine unstable mode, and no matter how
small that admixture is, $e^{\lambda_n t}$ eventually amplifies it to
dominate. We checked whether this was an artifact of interpolating
the eigenvector from the Chebyshev grid onto the SLAC evolution grid
by instead diagonalizing Eq.~\eqref{e:eigen-problem} directly on a
SLAC periodic grid ($N=1024$, same $L=160$ box as the evolution) and
transferring the result to the fine evolution grid by exact
band-limited (zero-padding) Fourier upsampling rather than spline
interpolation: both curves in
Fig.~\ref{f:SSPPOscillatoryContamination} agree closely throughout,
including the contamination onset time, so the effect is not specific
to either discretization -- it is ordinary floating-point roundoff in
the dense eigensolver, unavoidable whenever a problem's spectrum
contains both an exactly neutral mode and a genuinely growing one.

\subsection{\label{ss:VVPPstability}Specialization to VV-PP}

Take $q\to\infty$ (i.e.\ $g_s=0$) instead of $g_v=0$: $g_v=g/p>0$,
$g_p=g/r<0$ ($r<0$), the VV-PP case in which Fred's three-hump
examples (Sec.~\ref{ss:HumpStructure}) actually lie. Every result of
Secs.~\ref{ss:VKstability}--\ref{ss:BdG} applies unchanged with
$g_s\to0$: $H,M$ [Eqs.~\eqref{e:HH-def}--\eqref{e:MM-def}] simply lose
their $M_s$ terms,
\begin{align}
   H(x)
   &=
   \xi \, \partial_x + \sigma_3 \, m
   - \bigl[\, \omega_0 + g_v U^{\kappa} \,\bigr] - g_p \, W_{\kappa} \, \sigma_1
   \notag \\
   &\quad
   - g_v \kappa \, U^{\kappa-1} \, M_0
   + g_v U^{\kappa-1} \, M_v
   - g_p \kappa \, \lvert W\rvert^{\kappa-1} \, M_p \>,
   \notag \\
   M(x)
   &=
   - g_v \kappa \, U^{\kappa-1} \, M_0
   - g_v U^{\kappa-1} \, M_v
   - g_p \kappa \, \lvert W\rvert^{\kappa-1} \, M_p \>,
   \notag
\end{align}
and the two zero modes, Eqs.~\eqref{e:phase-zero-mode} and
\eqref{e:trans-zero-mode}, persist unchanged, since neither derivation
used $g_s\neq0$.

As a first, inexpensive check we apply the VK criterion of
Sec.~\ref{ss:VKstability} directly to the parameters underlying Fred's
Figs.~1--2 (Sec.~\ref{ss:HumpStructure}): $\kappa=1.05$, $p=1$,
$r=-0.95$. The existence boundary here (the $g_s=0$ analog of
Eq.~\eqref{e:omega-exist}, from $D(\gamma)=g_v-\lvert
g_p\rvert\sin^{\kappa+1}(2\gamma)\ge0$) is $\omega_{\rm
exist}\approx0.226\,m$ -- strikingly close to Fred's chosen
$\omega_0=0.23\,m$, confirming that his example sits just inside the
domain of existence. Evaluating $E_0/Q_0(\omega_0)$ numerically over
$\omega_{\rm exist}<\omega_0<m$ at these parameters, we find it
monotonically increasing throughout: the three-hump solitary wave of
Figs.~1--2, despite its nontrivial internal structure, is VK-stable.

\begin{figure}[t]
   \centering
   \includegraphics[width=\linewidth]{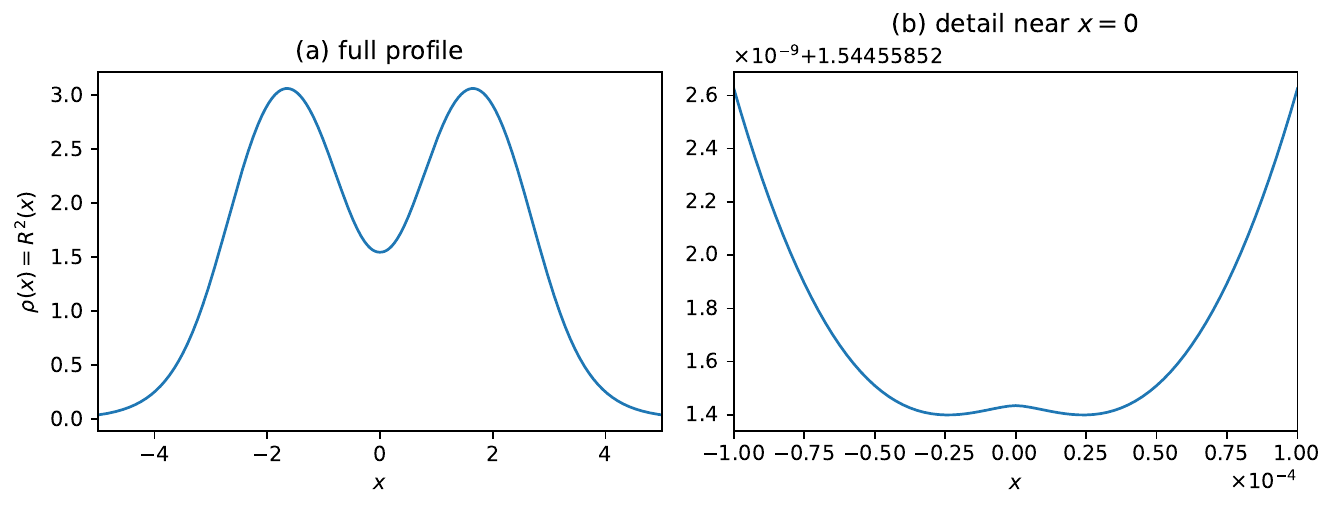}
   \caption{\label{f:VVPPFredReproduction}VV-PP, $\kappa=1.05$, $p=1$,
   $r=-0.95$, $\omega_0=0.23$ -- the parameters underlying Fred's Figs.
   1--2: (a) the full density profile $\rho(x)=R^2(x)$; (b) the
   detail near $x=0$, in 50-digit arbitrary precision, showing the
   genuine tiny central maximum and flanking minima.}
\end{figure}

As a further, independent check of this parameter identification,
Fig.~\ref{f:VVPPFredReproduction} evaluates our own closed-form
$R^2(x)$ [Eq.~\eqref{e:R2-y}] at exactly these parameters. Panel (a)
reproduces the two large side peaks ($\rho\approx3.06$ at
$x\approx\pm1.65$, matching Fred's ``magnitude around 3'' centered at
``$x=\pm1.6$'') and the central dip ($\rho\approx1.5446$, matching his
plotted $1.54456$ to five figures). Panel (b) -- computed in 50-digit
arbitrary precision, since the feature is otherwise buried in
double-precision roundoff until $x$ is resolved to $\sim10^{-10}$ --
shows the genuine fine structure underlying his Fig.~1 ``detail of
third hump'': a tiny local maximum exactly at $x=0$, flanked by two
minima near $x\approx\pm2.4\times10^{-5}$. The central maximum is
exactly the $\delta=1$ ($g_s=0$) prediction of
Sec.~\ref{ss:HumpStructure}, that $x=0$ is always a local maximum for
$\kappa>1$; the flanking minima are a finer-scale effect of the
pseudoscalar term's non-analytic $\lvert y\rvert^{\kappa+1}$
contribution, which is formally subleading as $y\to0$ but dominates
over an anomalously small range here, since the leading, smooth
$y^2$ coefficient of Eq.~\eqref{e:c2-general} is itself anomalously
small at these parameters.

As in the SS-PP case, scanning $\kappa$ at fixed $p=1,r=-0.95$ shows
$E_0/Q_0$ remains monotonic up to $\kappa\approx2$ and develops an
interior maximum -- signaling VK instability near $\omega_0\to m$ --
for $\kappa\gtrsim2.5$, the same qualitative pattern found for SS-PP
in Sec.~\ref{ss:SSPPstability}, though the critical $\kappa$ shifts
somewhat with $p,r$.

Since the VK criterion is only necessary, not sufficient, for linear
stability, we also directly diagonalized Eq.~\eqref{e:eigen-problem}
at Fred's own parameters ($\kappa=1.05$, $p=1$, $r=-0.95$,
$\omega_0=0.23$), using the Chebyshev method of
Sec.~\ref{ss:ChebyshevMethod}, across several resolutions
($N=250$--$400$, $L=10$--$15$). At every resolution a real
$\pm\lambda$ pair initially appears localized, but its value drifts by
tens of percent between grids ($0.29,0.23,0.34,0.28$ across the four
$(N,L)$ pairs checked) rather than converging -- precisely the
signature of a spurious box-discretization artifact rather than a
genuine mode, the same two-resolution test that caught a real missed
instability elsewhere in a companion paper
\cite{Cooper-Jacobs-Khare-2026}, here working in
the opposite direction. Once that artifact is rejected, the only
eigenvalues surviving between $N=300$ and $N=450$ are the two exact
zero modes and one additional genuine, purely oscillatory bound state
at $\pm0.497\,\rmi$ (real part $\sim10^{-16}$) -- the VV-PP
counterpart of the discrete mode found at
$\nu=\pm1.763\,\rmi$ for SS-PP in Sec.~\ref{ss:SSPPstability}, again
well inside the gap and again playing no role in the (absent)
instability here: no genuine growing mode remains. The three-hump
solitary wave of Figs.~1--2 is therefore linearly stable, not merely
VK-stable, at this parameter point.

\begin{figure}[t]
   \centering
   \includegraphics[width=\linewidth]{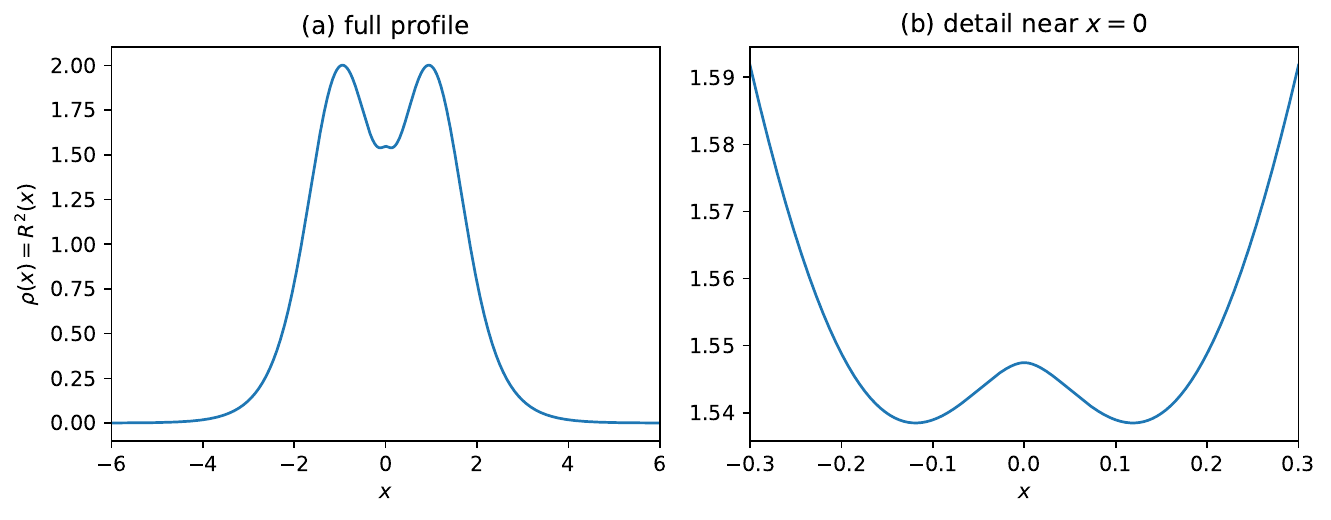}
   \caption{\label{f:VVPPFredReproductionKappa32}VV-PP, $\kappa=3/2$,
   $p=1$, $r=-0.95$, $\omega_0=0.23$ -- the parameters underlying
   Fred's Fig.~3: (a) the full density profile $\rho(x)=R^2(x)$; (b)
   detail near $x=0$, showing the modest but genuine central maximum
   and its flanking minima at $x\approx\pm0.119$, resolved here in
   ordinary double precision (contrast
   Fig.~\ref{f:VVPPFredReproduction}(b) at $\kappa=1.05$, where the
   analogous feature is $10^{5}$ times smaller and needs 50-digit
   precision to see at all).}
\end{figure}

Fred's own Fig.~3 is a second three-hump example at the same $p=1$,
$r=-0.95$, $\omega_0=0.23$ but $\kappa=3/2$ instead of $1.05$. The
existence boundary is now $\omega_{\rm exist}\approx0.201\,m$, a
wider margin than the $\kappa=1.05$ case, and correspondingly the
central feature no longer requires arbitrary precision to resolve:
Fig.~\ref{f:VVPPFredReproductionKappa32} evaluates our own closed-form
$\rho(x)$ directly in double precision, reproducing his three-hump
profile, with the two large maxima at $\rho(\pm0.944)\approx2.001$. At
$x=0$ there is a genuine, but still comparatively small, local
maximum $\rho(0)\approx1.5475$, flanked by minima at
$x\approx\pm0.119$ ($\rho\approx1.5385$) -- a height difference of
only $\sim0.009$, fully resolvable in ordinary double precision
[Fig.~\ref{f:VVPPFredReproductionKappa32}(b)] but still a far more
modest bump than the two dominant side peaks, consistent with the
$\delta=1$ prediction of Sec.~\ref{ss:HumpStructure} that $x=0$ is a
local maximum for $\kappa>1$, realized here over an ordinary, rather
than anomalously small, range of $x$, yet still a comparatively weak
one at these parameters.
The VK criterion again gives $dQ_0/d\omega_0=-65.8<0$ (stable), and
repeating the same multi-resolution diagonalization -- the same
non-converging spurious real pair appears at every single resolution
($0.45,0.34,0.55,0.42$ across four $(N,L)$ choices) and is rejected by
the same two-resolution test, leaving only the two zero modes between
$N=300$ and $N=450$, with no additional discrete oscillatory bound
state this time -- confirms this case is linearly stable as well.

\subsubsection{\label{ss:TimeEvolutionVVPP}Nonlinear time evolution}

\begin{figure}[t]
   \centering
   \includegraphics[width=\linewidth]{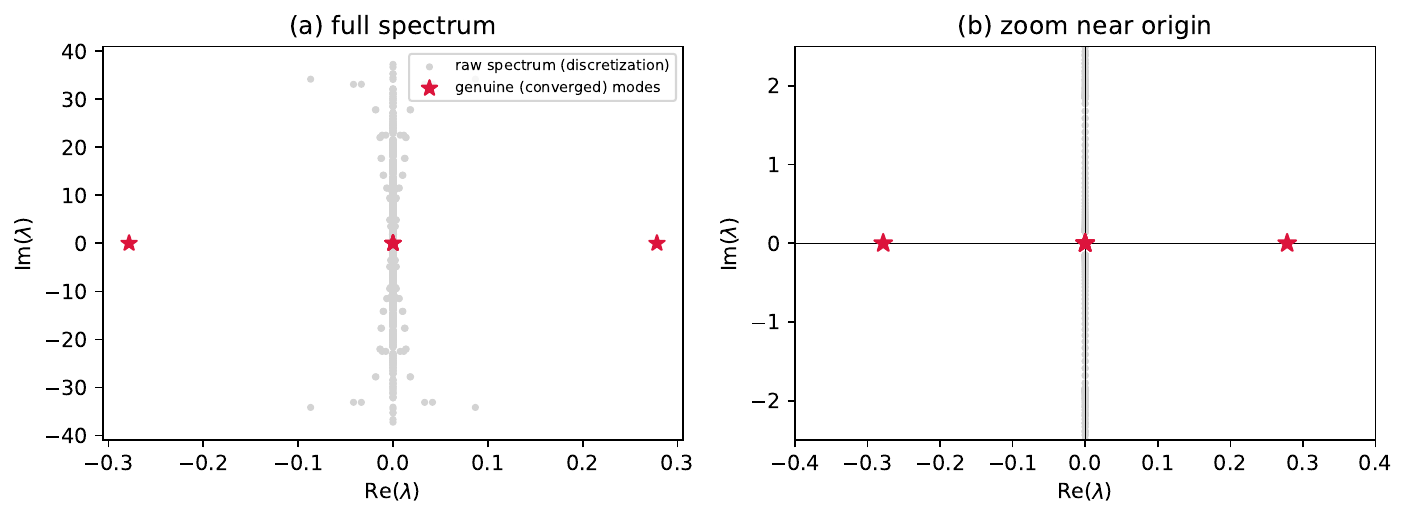}
   \caption{\label{f:VVPPSpectrum}The Bogoliubov-de~Gennes spectrum for
   the VV-PP example of Fig.~\ref{f:VVPPUnstableEigenmode}
   ($\kappa=3$, $p=1$, $r=-0.95$, $\omega_0=0.85$), at $N=300$,
   $L=8$, in the same format as Fig.~\ref{f:SSPPSpectrum}: (a) the
   full spectrum; (b) a zoom near the origin showing only the two
   zero modes and the unstable pair.}
\end{figure}

\begin{figure}[t]
   \centering
   \includegraphics[width=\linewidth]{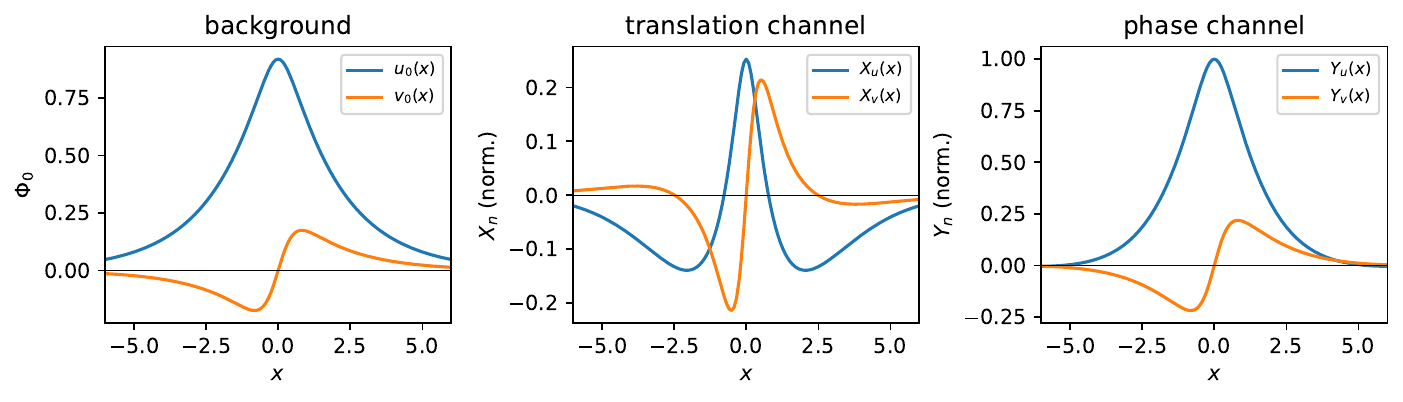}
   \caption{\label{f:VVPPUnstableEigenmode}VV-PP, $\kappa=3$, $p=1$,
   $r=-0.95$, $\omega_0=0.85$: background $\Phi_0=(u_0,v_0)$ (left) and
   the genuine unstable Bogoliubov-de~Gennes eigenmode $(X_n,Y_n)$
   (center, right; each normalized to unit peak height), with growth
   rate $\lambda_n=0.278$.}
\end{figure}

Scanning further in $\kappa$ at the same $p=1$, $r=-0.95$ turns up a
genuinely unstable VV-PP point well inside the regime anticipated in
Sec.~\ref{ss:VVPPstability} above: at $\kappa=3$, $\omega_0=0.85$, the
same multi-resolution diagonalization gives a growth rate
$\lambda_n=0.2783$, converged to five significant figures across five
independent $(N,L)$ choices and over three times larger than the
SS-PP growth rate of Sec.~\ref{ss:SSPPstability}. Relaxing the
growing-mode filter here finds no additional discrete oscillatory
bound state beyond the two zero modes and this unstable pair, unlike
the stable $\kappa=1.05$ case above: whether such a mode exists at all
depends on the parameters, not just on whether the soliton happens to
be stable or unstable.
Figure~\ref{f:VVPPSpectrum} shows this case's full spectrum in the
same format as Fig.~\ref{f:SSPPSpectrum}: only six eigenvalues survive
both filters here, not eight -- the two zero modes and the unstable
pair, with no third, oscillatory pair to find.
Figure~\ref{f:VVPPUnstableEigenmode} shows the background and this
unstable eigenmode, in the same format as Fig.~\ref{f:SSPPUnstableEigenmode}.

\begin{figure}[t]
   \centering
   \includegraphics[width=\linewidth]{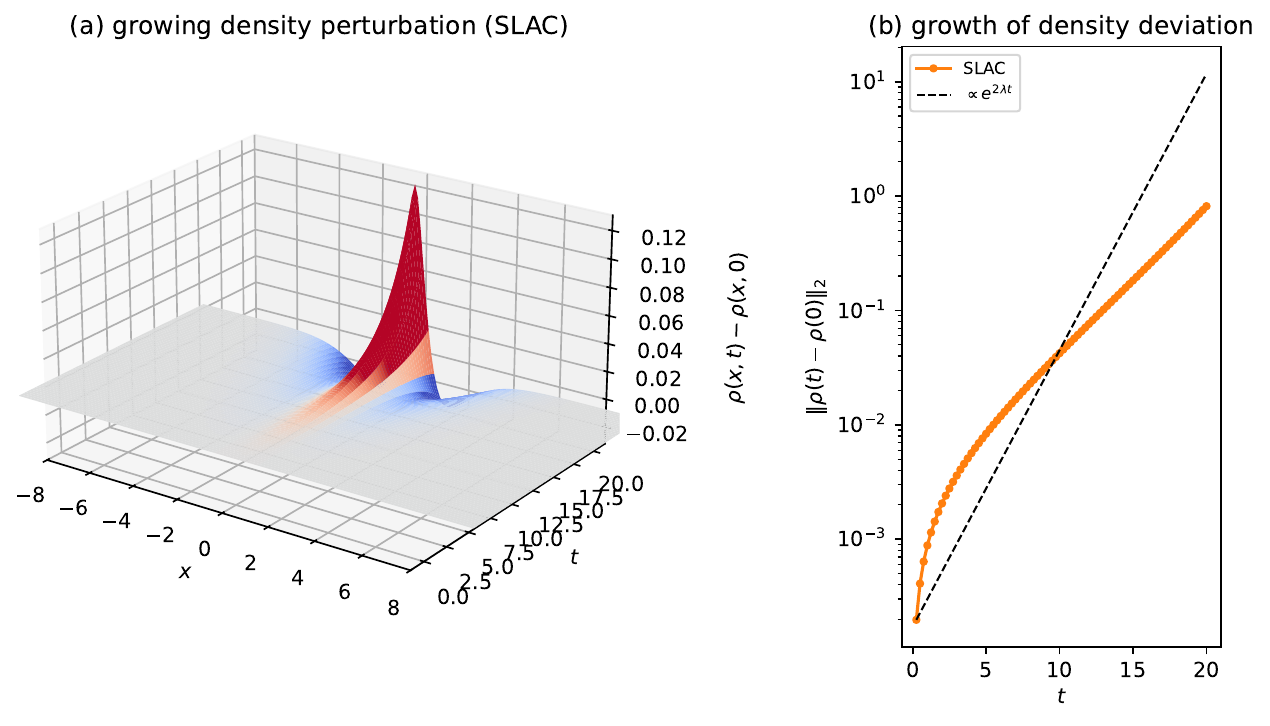}
   \caption{\label{f:VVPPTimeEvolution}Nonlinear time evolution of the
   perturbed VV-PP soliton of Fig.~\ref{f:VVPPUnstableEigenmode}: (a)
   the growing density perturbation $\rho(x,t)-\rho(x,0)$; (b) its
   $L^2$ norm, following $2\lambda_n$ (dashed) before departing from it
   as the perturbation grows into the nonlinear regime.}
\end{figure}

Repeating the nonlinear time evolution of Sec.~\ref{ss:TimeEvolution}
for this mode required one genuine extension: with $g_v\neq0$, the VV
term of Eq.~\eqref{e:EOM} no longer reduces to the static-only
$U^{\kappa}$ used there, since
$(\bar\Psi\gamma_\nu\Psi)(\bar\Psi\gamma^\nu\Psi)=U^2-J^2$ for a
genuinely complex, time-dependent $\Psi$ [cf.\
Eq.~\eqref{e:VVbracket}], with $J\equiv-2\,\mathrm{Im}(u^*v)$ the
analog of $W$ for the $\gamma^1$ current -- zero for the real static
$\Phi_0$ of Sec.~\ref{ss:ZerothOrder}, but generically nonzero once the
complex perturbation evolves. Writing $U_\kappa'\equiv
(U^2-J^2)^{(\kappa-1)/2}$, the correctly generalized equations of
motion are
\begin{align}
   \dot u &= \rmi\bigl[\,(g_sV_\kappa+g_vU_\kappa'U-m)\,u+g_pW_\kappa v-v'\,\bigr]
   + g_vU_\kappa'Jv \>,
   \notag \\
   \dot v &= \rmi\bigl[\,(m-g_sV_\kappa+g_vU_\kappa'U)\,v+g_pW_\kappa u+u'\,\bigr]
   - g_vU_\kappa'Ju \>,
   \notag
\end{align}
reducing to the $g_v=0$ equations of Sec.~\ref{ss:TimeEvolution}
exactly when $J=0$. We verified this both by confirming
$\dot\Psi=-\rmi\omega_0\Psi$ on $\Phi_0$ to machine precision
($\sim10^{-14}$, as in Sec.~\ref{ss:TimeEvolution}) and, independently,
by checking that its linearization about $\Phi_0$ along $X_n+\rmi Y_n$
reproduces $\lambda_n=0.27828$ to six figures via a direct numerical
Jacobian -- the same two checks used to validate the SS-PP equations.
Evolving from the same $10^{-3}\max|\Phi_0|$ perturbation as before,
charge is conserved to $1.6\times10^{-11}$ over the run, better even
than the SS-PP case. Figure~\ref{f:VVPPTimeEvolution} shows the
result: the same dip-and-peak growth pattern as Fig.~\ref{f:SSPPTimeEvolution},
developing faster (note the shorter time axis) and, consistent with
its larger $\lambda_n$, visibly entering the nonlinear regime by
$t\approx15$--$20$, where $\|\rho(t)-\rho(0)\|_2$ begins to fall below
the linear $e^{2\lambda_nt}$ reference.

A full account of the VV-PP soliton's reality domain, hump structure,
and a systematic map of Eq.~\eqref{e:eigen-problem}'s spectrum over
the full $(\kappa,p,r)$ family is left for future work.

%
% -----------------------------------------------------------------------
%
\section{\label{s:Conclusions}Conclusions}

We set out to determine the linear stability of solitary-wave
solutions of the nonlinear Dirac equation once a
pseudoscalar-pseudoscalar (PP) self-interaction is added alongside the
scalar-scalar (SS) and vector-vector (VV) terms studied elsewhere
\cite{Cooper-Jacobs-Khare-2026,Khare-Cooper-Dawson-Saxena-2026} -- a
three-coupling family for which neither the closed-form soliton's
stability properties nor a systematic Vakhitov-Kolokolov (VK) survey
had previously been worked out. We derived the closed-form soliton
solution, its reality domain, and its hump structure for general
couplings, then specialized to the two limits where a particular
choice of coupling makes the analysis sharpest: the SS-PP model
($g_v=0$) and the VV-PP model ($g_s=0$).

The VK criterion alone is only necessary, not sufficient, for linear
stability, so for both models we went beyond it: we derived the
Bogoliubov-de~Gennes (BdG) eigenvalue problem in general, solved it
numerically with a Chebyshev spectral discretization mapped onto the
whole real line, and filtered its output with a two-resolution
convergence test together with a spatial-localization criterion to
separate genuine eigenvalues from the discretized continuum. Every
growth rate reported as genuine was cross-validated a second way,
either against an independent SLAC-derivative lattice diagonalization
or by evolving the full nonlinear equation forward in time from a
perturbed initial condition and confirming that the observed growth or
oscillation matched the BdG prediction to several significant figures,
with charge conserved throughout to $10^{-11}$--$10^{-12}$ relative
precision.

For the VV-PP model, we confirmed that two representative three-hump
solitary waves ($\kappa=1.05$ and $\kappa=3/2$,
Sec.~\ref{ss:HumpStructure}) are not merely VK-stable but linearly
stable: every eigenvalue beyond the two zero modes has zero real part.
This directly answers, for these examples, a question left open by
prior nonlinear-time-evolution studies of multi-humped solitary
waves -- whether the instability those studies report is already
present at the linear level. We also located a genuinely unstable
VV-PP example ($\kappa=3$, $\omega_0=0.85$), extracted its unstable
eigenmode, and confirmed by nonlinear evolution that the perturbation
grows at the predicted rate $\lambda_n=0.2783$, more than three times
the analogous SS-PP rate. The SS-PP model shows the same qualitative
structure: VK instability appears once $\kappa\gtrsim2.2$ near
$\omega_0\to m$, and we confirmed a genuine unstable eigenmode
($\lambda_n=0.083$ at $\kappa=3$, $\omega_0=0.90$) directly, again with
the nonlinear evolution matching the predicted growth rate to four
significant figures across two independent numerical methods.

Beyond the translation and phase zero modes common to every case, a
broader, unfiltered scan of the spectrum turned up, in some but not
all of these examples, a further discrete mode: a genuinely stable,
purely oscillatory bound state sitting inside the spectral gap,
distinct in character from both the zero modes and the growing
instability. Exciting this mode and evolving the full nonlinear
equation confirms bounded oscillation at short times, as expected, but
reveals an instructive numerical limit at longer times: an
infinitesimal, unavoidable admixture of the system's own genuine
unstable mode -- present at the level of ordinary floating-point
roundoff, not a property of either discretization scheme we tested --
eventually grows to dominate. No numerically prepared initial
condition can stay exactly on a neutral mode forever when a genuine
instability exists nearby in the same spectrum.

Throughout, we benefited from close contact with a companion paper
that applies an independent, fully lattice-based version of this same
BdG analysis to the pure SS and pure VV models
\cite{Cooper-Jacobs-Khare-2026}; the agreement in spirit, together
with the complementary coverage -- SS and VV alone there, all three
couplings combined here -- suggests a natural road toward a single,
unified treatment of the whole SS+VV+PP family.

%
% -----------------------------------------------------------------------
%

% -----------------------------------------------------------------------
%
\begin{acknowledgments}
JFD acknowledges the assistance of Claude \cite{claude}, a large
language model developed by Anthropic, which served as a research
assistant and editor throughout, including substantial derivation and
verification work reported in this paper. All derivations were
independently verified against known limiting cases and by direct
numerical checks before being included in the text.
\end{acknowledgments}
%
% -----------------------------------------------------------------------
%
\appendix
%
% -----------------------------------------------------------------------
%
\section{\label{s:Notation}Notation}

In $1+1$ dimensions with $\rmd x^{\mu} = ( \rmd t, \rmd x)$, we use the
representations given in Refs.~\cite{PhysRevE.82.036604,Comech-2016} and
choose\footnote{We use the particle physics signature $(1,-1)$.}
\begin{equation}\label{e:gammaReps}
   \gamma^{0} = \sigma_3 =
   \begin{pmatrix} 1 & 0 \\ 0 & -1 \end{pmatrix}
   \,,\qquad
   \gamma^{1} = \rmi \sigma_1 =
   \rmi \begin{pmatrix} 0 & 1 \\ 1 & 0 \end{pmatrix}
   \,,
\end{equation}
where $\sigma_i$ are the Pauli matrices. $\gamma^5$ is given by:
\begin{equation}\label{e:gamma5-def}
   \gamma^5 = \gamma^{0} \gamma^{1} = -\sigma_2 = \rmi \, \xi \>,
   \qquad
   \gamma_5 = -\rmi \, \xi \>,
   \qquad
   \xi = \begin{pmatrix} 0 & 1 \\ -1 & 0 \end{pmatrix} \,.
\end{equation}
The charge-conjugation matrix $\calC$ of the $\bbZ_2^{\calC}$ symmetry
$\Psi\rightarrow\calC\,\Psi$ in Sec.~\ref{s:Model} is fixed by
requiring this map to preserve the free Dirac equation, i.e.
$\calC^{-1}\gamma^{\mu}\calC=-(\gamma^{\mu})^{\ast}$ for each $\mu$.
Since $\gamma^0$ is real, $\calC$ must anticommute with $\gamma^0$; since
$\gamma^1$ is purely imaginary, $\calC$ must commute with $\gamma^1$. The
unique Pauli matrix satisfying both is
\begin{equation}\label{e:C-def}
   \calC = \sigma_1 \, \calK \>,
\end{equation}
where $\calK$ is the antilinear complex-conjugation operator,
$\calK\Psi\equiv\Psi^{\ast}$. $\calC$ is real, Hermitian, and satisfies
$\calC^2=1$, so that $\Psi\rightarrow\calC\,\Psi=\sigma_1\,\Psi^{\ast}$
is a genuine order-two symmetry. Since
$\gamma^5=\gamma^0\gamma^1$ with $\calC$ anticommuting with $\gamma^0$ and
commuting with $\gamma^1$, it follows that
\begin{equation}\label{e:C-gamma5-anticommute}
   \{ \gamma^5, \, \calC \} = 0 \>.
\end{equation}
The gamma matrices then obey the
anti-commutation relation, $\{ \gamma^{\mu}, \, \gamma^{\nu} \} = 2
g^{\mu\nu}$ as required. Using these representations, the bracket
terms appearing in Eq.~\eqref{e:VSSVVPP} evaluate to
\begin{subequations}\label{e:BracketTerms}
\begin{align}
   (\bar\Psi \Psi)(\bar\Psi \Psi) &= (\Psi^{\dagger} \sigma_3 \Psi)^2 \>,
   \label{e:SSbracket} \\
   (\bar\Psi \gamma_{\mu} \Psi)(\bar\Psi \gamma^{\mu} \Psi)
   &=
   (\Psi^{\dagger} \Psi)^2 - (\Psi^{\dagger} \sigma_2 \Psi)^2
   =
   (\Psi^{\dagger} \Psi)^2 + (\Psi^{\dagger} \xi \Psi)^2 \>,
   \label{e:VVbracket} \\
   (\bar\Psi \gamma_{5} \Psi)(\bar\Psi \gamma^{5} \Psi)
   &=
   (\Psi^{\dagger} \sigma_3 \xi \Psi)^2 = (\Psi^{\dagger} \sigma_1 \Psi)^2 \>,
   \label{e:PPbracket}
\end{align}
\end{subequations}
where the last equality uses $\sigma_3 \xi = \sigma_1$; the PP bracket
is thus manifestly real and non-negative for any $\Psi$.

For the linear stability analysis of Sec.~\ref{s:Stability}, we will
also need the following matrices, built from outer products of the
background solution $\Phi_0=(u_0,v_0)^T=R(\cos\theta,\sin\theta)^T$
[Eq.~\eqref{e:Phi0-solution-form}]:
\begin{subequations}\label{e:AllMatrices}
\begin{align}
   M_0
   &=
   \Phi_0 \, \Phi_0^{T}
   =
   \begin{pmatrix} u_0 & v_0 \end{pmatrix}
   \begin{pmatrix} u_0 \\ v_0 \end{pmatrix}
   =
   \begin{pmatrix} u_0^2 & u_0v_0 \\ u_0v_0 & v_0^2 \end{pmatrix}
   \notag \\
   &=
   R^2 \begin{pmatrix} \cos^2(\theta) & \sin(\theta)\cos(\theta) \\
             \sin(\theta)\cos(\theta) & \sin^2(\theta) \end{pmatrix} \>,
   \label{e:M0-def} \\
   M_{s}
   &=
   \sigma_3 \Phi_0 \, \Phi_0^{T} \sigma_3
   =
   \begin{pmatrix} u_0 & -v_0 \end{pmatrix}
   \begin{pmatrix} u_0 \\ -v_0 \end{pmatrix}
   =
   \begin{pmatrix} u_0^2 & -u_0v_0 \\ -u_0v_0 & v_0^2 \end{pmatrix}
   \notag \\
   &=
   R^2 \begin{pmatrix} \cos^2(\theta) & -\sin(\theta)\cos(\theta) \\
            -\sin(\theta)\cos(\theta) & \sin^2(\theta) \end{pmatrix} \>,
   \label{e:Ms-def} \\
   M_{p}
   &=
   \sigma_1 \Phi_0 \, \Phi_0^{T} \sigma_1
   =
   \begin{pmatrix} v_0 & u_0 \end{pmatrix}
   \begin{pmatrix} v_0 \\ u_0 \end{pmatrix}
   =
   \begin{pmatrix} v_0^2 & u_0v_0 \\ u_0v_0 & u_0^2 \end{pmatrix}
   \notag \\
   &=
   R^2 \begin{pmatrix} \sin^2(\theta) & \sin(\theta)\cos(\theta) \\
             \sin(\theta)\cos(\theta) & \cos^2(\theta) \end{pmatrix} \>,
   \label{e:Mp-def} \\
   M_{v}
   &=
   -\xi \Phi_0 \, \Phi_0^{T} \xi
   =
   \begin{pmatrix} -v_0 & u_0 \end{pmatrix}
   \begin{pmatrix} -v_0 \\ u_0 \end{pmatrix}
   =
   \begin{pmatrix} v_0^2 & -u_0v_0 \\ -u_0v_0 & u_0^2 \end{pmatrix}
   \notag \\
   &=
   R^2 \begin{pmatrix} \sin^2(\theta) & -\sin(\theta)\cos(\theta) \\
            -\sin(\theta)\cos(\theta) & \cos^2(\theta) \end{pmatrix} \>,
   \label{e:Mv-def}
\end{align}
\end{subequations}
with $\xi$ as in Eq.~\eqref{e:gamma5-def}. Note $M_v=-\xi M_0\xi$
rather than $+\xi M_0\xi$: since $\xi$ is antisymmetric (unlike the
symmetric $\sigma_1,\sigma_3$ used in $M_p,M_s$), the sign of the
outer $\xi$-dressing matters. The Pauli matrices used throughout are
\begin{equation}\label{e:PM.Pauli}
   \sigma_1 = \begin{pmatrix} 0 & 1 \\ 1 & 0 \end{pmatrix}
   \,,\qquad
   \sigma_2 = \begin{pmatrix} 0 & -\rmi \\ \rmi & 0 \end{pmatrix}
   \,,\qquad
   \sigma_3 = \begin{pmatrix} 1 & 0 \\ 0 & -1 \end{pmatrix} \,.
\end{equation}

\bibliographystyle{apsrev4-2}
\nocite{apsrev42Control}
\bibliography{cooper-SSPP.bib}
\end{document}